\documentclass[acmsmall, nonacm, screen]{acmart}

\usepackage [ruled, linesnumbered]{algorithm2e}
\usepackage{url}
\usepackage{mathtools}
\usepackage{multirow}
\usepackage{makecell}
\usepackage[encapsulated]{CJK}
\usepackage[table, dvipsnames]{xcolor}
\usepackage{adjustbox}
\usepackage{subcaption}
\usepackage{threeparttable}
\usepackage{pifont}

\definecolor{textred}{RGB}{231, 29, 52}
\definecolor{textblue}{RGB}{1, 146, 213}
\definecolor{codegray}{gray}{0.9}

\newcommand{\code}[1]{``{\texttt{\small{#1}}}''}

\newcommand{\codeWoQuot}[1]{{\texttt{\small{#1}}}}

\setcopyright{acmlicensed}

\title{RulER: Automated Rule-Based Semantic Error Localization and Repair for Code Translation}

\begin{document}

\author{Shuo Jin}
\authornote{Equal contribution and co-first authors.}
\email{imjinshuo@whu.edu.cn}
\affiliation{\institution{School of Computer Science, Wuhan University}\country{China}}

\author{Songqiang Chen}
\authornotemark[1]
\email{i9s.chen@connect.ust.hk}
\affiliation{\institution{{The Hong Kong University of Science and Technology}\country{China}}}

\author{Xiaoyuan Xie}
\authornote{Corresponding author.}
\email{xxie@whu.edu.cn}
\affiliation{\institution{School of Computer Science, Wuhan University}\country{China}}

\author{Shing-Chi Cheung}
\email{scc@cse.ust.hk}
\affiliation{\institution{{The Hong Kong University of Science and Technology}\country{China}}}

\begin{abstract} 
Automated code translation aims to convert programs between different programming languages while maintaining their functionality.
Due to the imperfections of code translation models, the generated translations may contain errors that compromise their reliability. Existing automated debugging methods for code translation rely on code alignments and repair patch templates to locate and fix erroneous translations. 
However, existing methods lack reliable references to construct code alignments and design repair patch templates, which significantly impacts their localization accuracy and repair effectiveness.
To address these limitations, we reintroduce code translation rules and propose a rule-based debugging method for code translation, called RulER. RulER automatically derives large-scale code translation rules from correct translations generated by large language models, enabling the efficient collection of diverse translation rules. In addition, RulER dynamically combines the existing rules on expandable nodes like expressions and tokens to further adaptively align more statements.
These rules capture clear and detailed structural correspondences between source and target programming languages. Therefore, they can serve as reliable and reusable references for code alignment and repair template design, enabling RulER to locate and fix translation errors effectively.
Our evaluation of RulER on Java-to-C++ and Python-to-C++ translations produced by four code translation models demonstrates that RulER outperforms state-of-the-art methods, BatFix and TransMap. Our experimental results show that RulER outperformed the best baseline by 20\% and 272\% in terms of error localization rates and repair success rates, respectively.
RulER exhibits superior repair performance compared to directly prompting LLMs for patch generation, demonstrating a promising methodology for extracting and leveraging coding knowledge from LLMs.
\end{abstract}

\keywords{Code translation, Error localization, Large language model, Rule-based repair}

\maketitle

\begin{CJK*}{UTF8}{gbsn}
\section{Introduction}

Code translation aims to convert a program written in one Programming Language (PL) to another while preserving the original logic and functionality. It shows wide applications, e.g., modernizing legacy software~\cite{modernization-1, modernization-2, modernization-3, modernization-4, modernization-5}, enabling multi-platform software development~\cite{multiplatform-1, multiplatform-2, multiplatform-3}, and facilitating model training to achieve better code generation performance~\cite{modeltrain-1, modeltrain-2, modeltrain-3, modeltrain-4}. Therefore, automated code translation is crucial for accelerating software migration, reducing development costs, and improving development efficiency. Initially, several rule-based code transpilers were developed~\cite{CxGo, C2Rust, JavaToCSharp}. However, such rule-based tools were criticized for generating translations with low readability and maintainability~\cite{TransCoder}. Later, various learning-based methods have been proposed. These methods include designing Deep Learning (DL) models specialized for code translation~\cite{Tree-to-tree, TransCoder, TransCoder-ST, IR-augmented-translation-method, BabelTower, ADELT, Attention-Need, VIM-PT, CoDist, SDA-Trans}, and prompting Large Language Models (LLMs) to generate translations directly~\cite{InterTrans, UniTrans, Low-resource-trans, LLM-LIFT}. These methods have significantly improved the readability and accuracy of translated code and have gradually become mainstream in this field.

Despite the advancements in learning-based code translation methods, the translations generated by these approaches may still be incorrect.
Many incorrect translations have correct syntax yet suffer from semantic errors~\cite{Code-Translation-Survey2}. The semantic errors manifest as discrepancies in logic or functionality between the target translation and the source program, such as incorrect if conditions, resulting in different execution outcomes between the two programs. 
To correct such buggy translations, two automated debugging methods have been proposed to locate and repair semantic errors in code translation~\cite{BatFix, TransMap}. 
Both methods operate based on the functionally equivalent statements aligned between the source and target programs. 
By comparing the runtime information of aligned code snippets under identical test inputs, these methods identify discrepancies and pinpoint the incorrectly translated code snippets causing the semantic errors. 
Erroneous translation statements can be further repaired by referencing their corresponding aligned source statements. 

However, existing state-of-the-art structural graph-driven and LLM-based methods, BatFix \cite{BatFix} and TransMap \cite{TransMap}, exhibit limitations in locating and repairing semantic errors due to the lack of reliable references for building code alignments and designing repair patch templates.
Specifically, BatFix aligns the code statements based on the Control Flow Graphs (CFGs) of the source and target programs. 
However, CFGs fail to serve as robust alignment references once the source and target programs exhibit different CFG structures.
TransMap uses a one-shot example to prompt the LLM to generate line-to-line code alignments. However, this example fails to provide references for identifying code alignments that are not represented within the example. Consequently, when dealing with translations that differ from the programs in the example, TransMap may produce incorrect or incomplete alignments.
Besides, these methods also have limitations in error repair. TransMap does not support automated correction of translation errors. While BatFix attempts to fix the located translation errors by replacing them with a patch template abstracted from the source program, such templates often follow the syntax of the source PL and fail to fit the target PL, thereby hindering an effective fix. 
In summary, while both methods realize the automated debugging of code translations, the lack of reliable references for building code alignments and designing repair templates limits their effectiveness in mitigating translation errors.

In this work, we reintroduce code translation rules, leveraging them as
references for code alignment and repair patch generation. Code translation rules define transformation relationships between different PLs while preserving equivalent functionality. They show great potential to address the limitations of existing debugging methods. Specifically, for code alignment, translation rules can capture mappings between significantly divergent code structures, enabling more accurate alignment for translation pairs with different CFG structures than CFG-based BatFix. Besides, compared to the one-shot example used by TransMap for guiding code alignment, we can flexibly select suitable translation rules for different translation pairs. These selected rules can provide tailored guidance for each specific translation pair for code alignment. 
Furthermore, for patch generation, translation rules enable the derivation of patch generation templates that better align with the syntactic requirements of the target PL. This enables the production of more effective repair patches than those of BatFix, which are generated based on the source code structure directly.

Following this idea, we propose a rule-based debugging method called RulER.
To mitigate the high cost of manually designing code translation rules, RulER extracts translation rules from correct code translations automatically. With the advancement of LLMs, LLMs can work as automated code translators to produce numerous correct code translation results that pass the test validation. Leveraging such diverse correct LLM-generated code translations, RulER employs a removal-based strategy to extract large-scale statement-level code translation pairs. These translation pairs provide a solid and diverse foundation for RulER to derive translation rules applicable to various code statements across different PLs.
In addition, based on the mined rules, RulER further dynamically synthesizes new rules by combining the existing ones on expandable nodes like expressions and tokens to apply to more statements with varied expression structures.
These mined and synthesized rules then serve as references to guide the construction of code alignments between the source program and its translation. With the alignments, RulER locates semantic errors by detecting runtime divergences between the aligned codes. 
RulER also synthesizes more accurate repair templates to facilitate error translation repair by leveraging the expected translation structures derived from the translation rules.

To evaluate RulER, we reuse the dataset from \citet{BatFix} and extend it to incorporate translations generated by a recently developed LLM. 
The dataset includes 553 erroneous code translations generated by four representative code translation models from Java and Python to C++.
Evaluation results demonstrate that RulER can provide translation rules applicable to 92.6\% distinct source statements in the dataset, where the dynamic synthesis of expanded rules contributes to 54.2\%. Based on these rules, RulER achieves an average F1 score of 96.1\% in building code alignments, outperforming BatFix's 74.1\% and TransMap's 82.9\%. The better code alignments enable RulER to achieve an average error localization success rate of 77.6\%, with relative improvements of 65\% and 20\% over BatFix and TransMap, respectively. Furthermore, RulER generates more accurate patches than BatFix, achieving a 272\% improvement in repair success rate. Particularly for translations from Python to C++, whose syntax and usages differ fairly significantly, RulER achieves a remarkable improvement of 784\% over BatFix in repair success rate. We also compare RulER against directly prompting an LLM for patch generation. RulER still demonstrates a relative improvement of 56\% over the LLM in repair success rate, highlighting the effectiveness of translation rules mined from LLMs in generating correct translation patches.

To summarize, this paper addresses the limitations of existing code translation debugging methods in locating and repairing semantic errors by applying massive automatically mined translation rules. 
Our contributions are outlined below:

\begin{itemize}

\item We revisit the role of translation rules and reintroduce them into automated code translation in a new capacity, serving as a relatively rigorous reference to refine the erroneous translations generated by learning-based code translation models.

\item We design and implement a rule-based end-to-end debugging method for buggy code translation, called RulER. RulER employs a removal-based approach to automatically extract code translation rules from LLM-generated translations and dynamically synthesizes new rules by combining existing ones. It then references these rules to establish code alignments for error localization and further applies them to derive the templates to generate repair patches for fixing erroneous translations.

\item With RulER, we have constructed and open-sourced 30,182 code translation rules across Java, Python, and C++. These rules are proven to effectively cover 92.6\% of source code statements in the evaluation dataset.

\item Experimental results demonstrate that RulER outperforms existing baseline methods in code alignment construction, error localization, and repair patch generation. Based on translation rules, RulER more effectively locates and repairs semantic errors in code translations compared with baseline methods, with relative improvements over the best baseline reaching 20\% and 272\% in error localization rates and repair success rates, respectively.

\end{itemize}

The remaining paper is organized as follows. In Section~\ref{sec:background}, we introduce the task of code translation and existing methods for locating and repairing semantic errors in code translations. Section~\ref{sec:motivation} presents the limitations of existing methods in building code alignments and designing patch generation templates, as well as our idea to address these limitations. In Section~\ref{sec:methodology}, we present our rule-based method RulER and introduce its four modular components in detail. Section~\ref{sec:expsetup} outlines the experimental setup, and Section~\ref{sec:expresult} presents the experiment results and analysis for the research questions. In Section~\ref{sec:discussion}, we revisit the role of code translation rules in automated code translation and present the types of semantic errors that RulER can fix. Section~\ref{sec:threatstovalidity} discusses the threats to validity of this work. In Section~\ref{sec:relatedwork}, we review related works on fault localization and automated program repair. Finally, we conclude the paper and outline directions for future study in Section~\ref{sec:conclusion}.

\section{Background}
\label{sec:background}

\subsection{Code Translation}

Given a program $S$ written in the source PL $\mathcal{L}_S$, code translation aims to produce a program $T$ in the target PL $\mathcal{L}_T$, such that $T$ performs the same functionality as $S$.
To achieve automated code translation, early efforts focused on developing \textit{rule-based code transpilers}, such as JavaToCSharp~\cite{JavaToCSharp}, py2many~\cite{py2many}, C2Rust~\cite{C2Rust}, and CxGo~\cite{CxGo}. These tools rely on manually crafted rewrite rules to translate the source programs to the target PL. Despite their effectiveness in generating correct translations, practitioners identify notable limitations of them. First, designing rules is labor-intensive and requires expertise in multiple PLs~\cite{TransCoder}. Second, translations generated by rules often suffer from poor readability and maintainability~\cite{IR-augmented-translation-method}.

Later, learning-based code translation methods became mainstream due to their efficiency and generalizability~\cite{Tree-to-tree, TransCoder, TransCoder-ST, IR-augmented-translation-method, BabelTower, ADELT, Attention-Need, CoDist}. Inspired by advancements in machine translation techniques, DL models trained on large-scale monolingual code corpora show promising capabilities in generating human-readable code translations~\cite{IR-augmented-translation-method}. With the emergence of powerful LLMs, recent studies have explored leveraging LLMs for code translation~\cite{InterTrans, UniTrans, Low-resource-trans, LLM-LIFT, SDA-Trans}.
However, learning-based code translators are not yet fully reliable and may produce erroneous translations~\cite{Code-Translation-Survey1, Code-Translation-Survey2, Code-Translation-Survey3}, such as inconsistent if conditions, incorrect API usage, and incorrect data types.

\subsection{Semantic Errors in Code Translation}

Code translators may yield incorrect translations, particularly the popular learning-based methods~\cite{Code-Translation-Survey2}. Incorrect translations may include syntax and semantic errors. 
\textit{Syntax errors} in code translations, which cause the translation to violate the grammar of the targeted PL, are easy to identify and fix, as in general code generation scenarios. 
Meanwhile, \textit{semantic errors}, which manifest as functional deviation from the source programs, are much more challenging to locate and fix.

\begin{figure*}[t]
\centering
\includegraphics[width=1.0\linewidth]{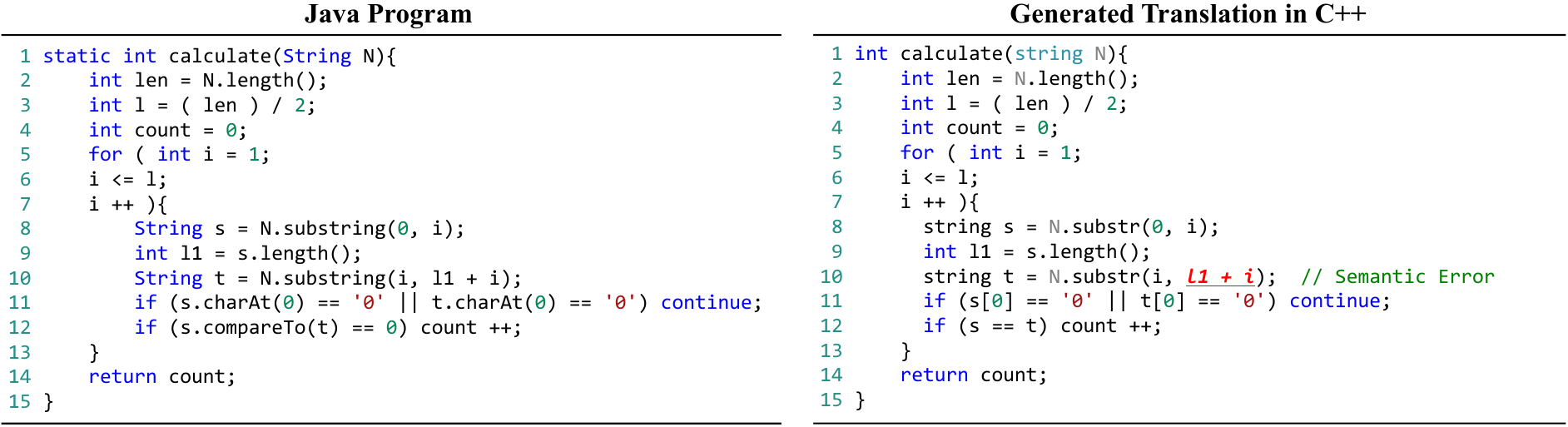}
\caption{Semantic error in code translation generated by TransCoder.}
\label{SemanticError}
\end{figure*}

In code translation, semantic errors arise when the translated code deviates from the intended logic or functionality of the source code, such as incorrect logical conditions or improper usage of APIs. Figure~\ref{SemanticError} demonstrates a semantic error produced by TransCoder when translating the Java expression \code{N.substring(i, l1+i);} into the C++ expression \code{N.substr(i, l1+i);}. While the generated C++ expression is syntactically valid, it includes a semantic mismatch due to the translator ignoring the differences in the usage of Java API \code{substring()} and C++ API \code{substr()}.
Specifically, \code{N.substring(i, l1+i);} in Java extracts a substring of the string \codeWoQuot{N} starting at index \codeWoQuot{i} and ending at index \codeWoQuot{l1+i}. In contrast, \code{N.substr(i, l1+i);} in C++ extracts a substring of \codeWoQuot{N} starting at index \codeWoQuot{i} with a length of \codeWoQuot{l1+i}. Actually, the correct translation should be \code{N.substr(i, l1);} to perform equivalent functionality.

Semantic errors are challenging to locate and fix as they are more subtle and involve discrepancies in program behavior~\cite{TransMap}. Identifying the fix location of the semantic errors in $T$ requires developers proficient in both $\mathcal{L}_S$ and $\mathcal{L}_T$ to meticulously compare $T$ with $S$, often on a line-by-line basis. The required effort increases as the complexity of the translation grows~\cite{BatFix, TransMap}.

\subsection{Automated Error Localization and Repair for Code Translation}

Existing automated debugging methods rely on code alignment between source and target programs to locate and resolve semantic errors.
Specifically, code alignment establishes the correspondences between statements in source program $S$ and target program $T$ that implement equivalent functionality. 
Semantic errors can be located by gradually comparing the runtime information of aligned codes.
If any divergence is detected in the execution traces or variable values between the aligned codes, a semantic error is located.
Existing methods rely on the structural similarity between two programs or turn to artificial intelligence to build code alignments. 
Specifically, BatFix \cite{BatFix} aligns statements between $S$ and $T$ by analyzing their CFGs to align the code blocks with similar CFG structures. TransMap \cite{TransMap} takes a different approach by prompting an LLM to generate line-to-line code alignments between $S$ and $T$ directly. 

After locating the semantic errors in translation, BatFix further generates repair patches to fix the erroneous translation. 
Specifically, given an erroneous statement $t$ in $T$, BatFix locates the corresponding statement $s$ in $S$ according to the code alignments. BatFix assumes that $s$, despite being written in a different PL from the translation, provides a reference solution for repairing the erroneous translation $t$. Therefore, BatFix designs a repair template based on $s$, which is an incomplete code sketch representing the structure of repair patches. Then, BatFix replaces the placeholders in the template with specific variables, constants, and function names to generate concrete repair patches, which are used to replace the erroneous $t$ in $T$.

\section{Preliminary}
\label{sec:motivation}

\subsection{Limitations in Locating and Fixing Semantic Errors}

While the state-of-the-art methods, BatFix and TransMap, can locate and fix semantic errors in some code translations, their localization accuracy and repair effectiveness are constrained by the quality of the code alignments and repair patch templates they generate.

\subsubsection{Limitations in Building Code Alignments for Locating Semantic Errors}

\begin{figure*}[t]
\centering
\includegraphics[width=1.0\linewidth]{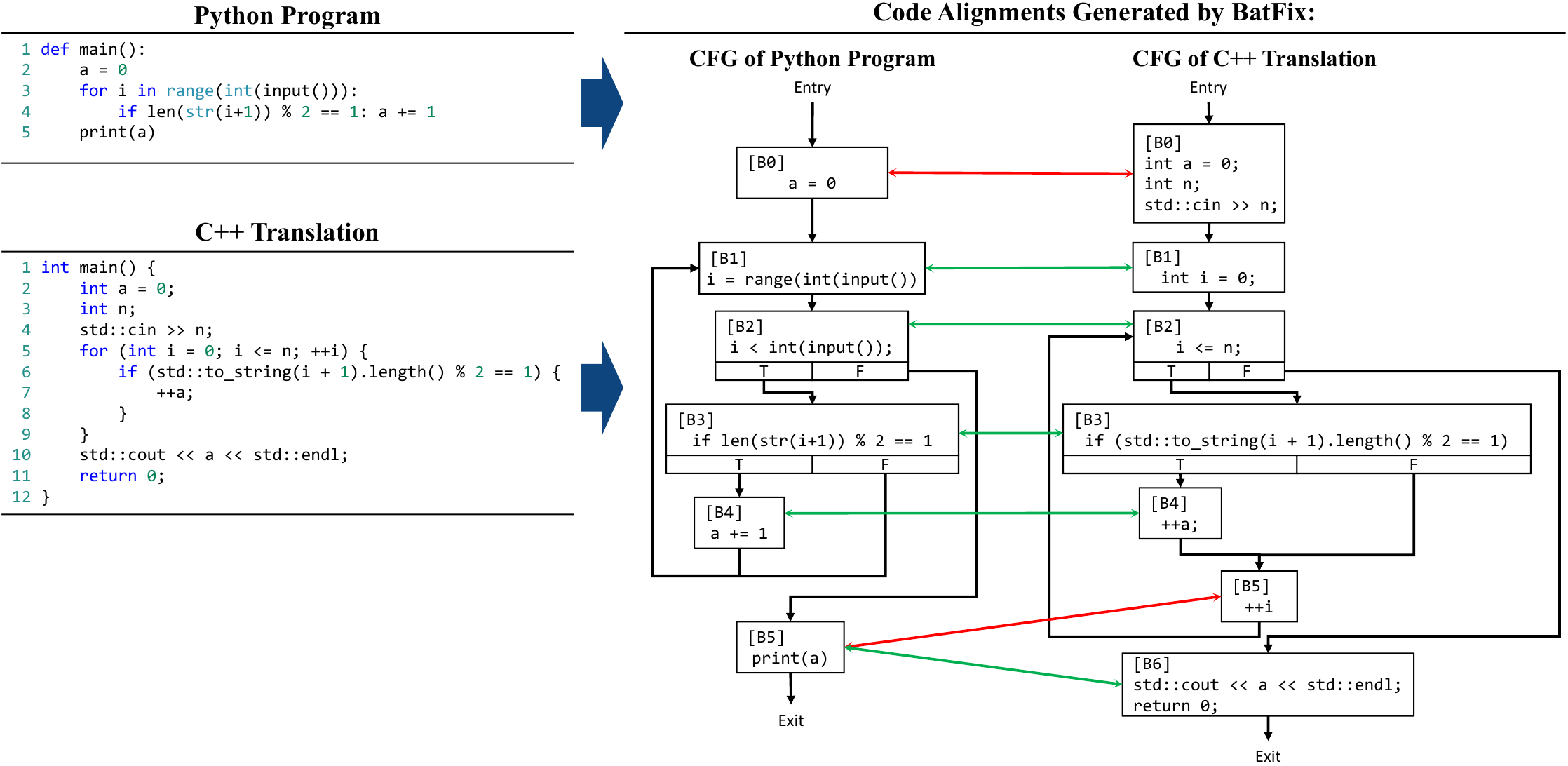}
\caption{Alignments built by BatFix between a translation pair with different CFG structures. The red arrows indicate the incorrect alignments.}
\label{Model-Example1}
\end{figure*}

Code alignment serves as the foundation for detecting runtime divergences between source and target programs, which is critical for pinpointing the locations of semantic errors accurately. However, both the existing debugging methods, BatFix and TransMap, lack reliable references for building code alignments, which hinders their accuracy in locating the semantic errors.

Specifically, BatFix parses the CFGs of the source and target programs as references to align code statements between them. However, programs that achieve the same functionality but are written in different PLs can have different CFG structures. The differences render CFGs unreliable as references, leading BatFix to produce incorrect code alignments. For example, as shown in Figure~\ref{Model-Example1}, the statement \code{for i in range(int(input())):} in a Python program is translated into C++ as \code{int n; std::cin>>n; for(int i=0; i<n; ++i)}. Although the translation implements equivalent functionality, its CFG structure is different from the source code: (1) the first block of Python CFG ([B0]: \code{a=0}) does not correspond to the first block of C++ CFG ([B0]: \code{int a=0; int n; std::cin>>n;}). (2) introducing an additional block ([B5]: \code{++i}) in the C++ CFG. According to their different CFGs, BatFix constructs two incorrect code alignments as indicated by the red arrows, which lead BatFix to interpret the location of the semantic error incorrectly. Specifically, when comparing the execution paths of both programs step by step, BatFix reports a trace divergence after executing the two [B4] blocks, where the Python code proceeds to [B1] and the C++ code steps to [B5]. Although Python's [B1] and C++'s [B5] blocks achieve equivalent functionality of updating the value of \codeWoQuot{i}, there is no alignment relationship between them based on the CFGs. As a result, BatFix identifies the semantic error at [B4] in the C++ translation mistakenly.

\begin{figure*}[t]
\centering
\includegraphics[width=0.86\linewidth]{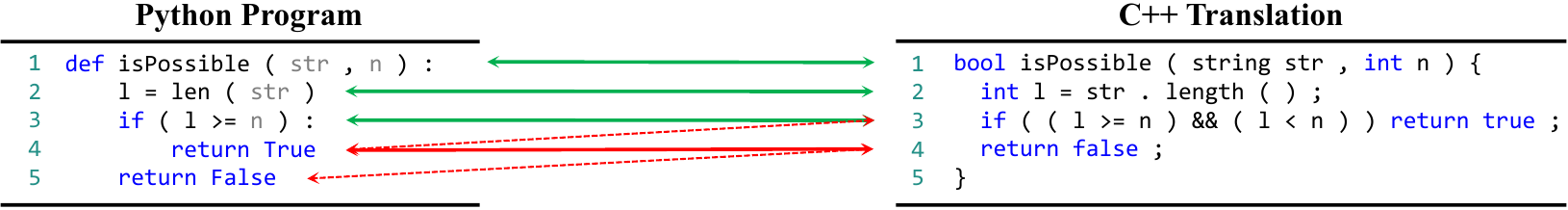}
\caption{Line-to-line code alignments generated by TransMap. The solid red arrow indicates incorrect alignment, while the dashed red arrows represent alignments missed by TransMap.}
\label{Model-Example2}
\end{figure*}

TransMap prompts an LLM to generate line-to-line alignments between the source program and its target translation. However, TransMap uses only a one-shot example to prompt the LLM for generating code alignments across various translation pairs. The example is a short source code snippet and its translation, along with manually annotated alignment relationships between them. 
While this example provides the output format for generating code alignment, it fails to provide references for identifying alignments beyond those annotated in the example.
As a result, for translation pairs that differ from the programs in the example, TransMap may produce incorrect or incomplete alignment results due to the lack of references.
For example, as shown in Figure~\ref{Model-Example2}, the statement \code{return True} in the source Python program and the statement \code{return false} in the C++ translation, neither of which appears in the one-shot example, are incorrectly aligned by TransMap, as indicated by the solid red arrow. In addition, two expected alignment relationships, as shown by the dashed red arrows, are missed by TransMap.
As a result, these incorrect alignments prevent TransMap from identifying an execution trace divergence following their respective \code{if} statements, where the Python program moves to \code{return True} but the C++ program steps to \code{return false}. Since TransMap incorrectly aligns the two statements, it ultimately fails to identify the semantic error within the preceding \code{if} statement in the C++ translation.

\subsubsection{Limitations in Designing Templates for Patch Generation}

\begin{figure*}[t]
\centering
\includegraphics[width=1.0\linewidth]{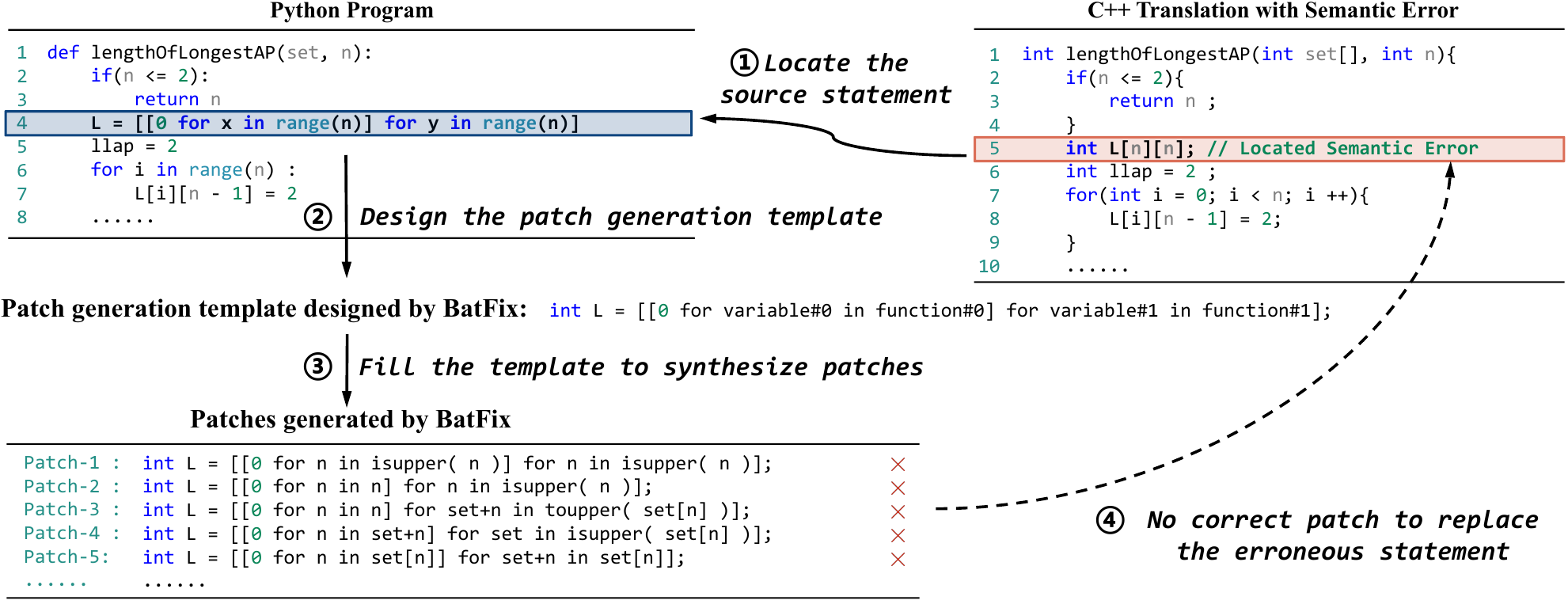}
\caption{A running example of BatFix failing to generate a correct patch generation template for an erroneous translation.}
\label{Model-Example3}
\end{figure*}

TransMap does not provide functionality for fixing translation errors automatically, whereas BatFix attempts to repair the erroneous translations by referencing their corresponding source code.
However, BatFix struggles with designing correct code templates to generate effective repair patches. 
Specifically, for a suspicious translated statement $t$, BatFix references the corresponding statement $s$ from the source program to design a repair template. 
The template replaces the identified buggy snippet in translation outputs and fills in its placeholders with appropriate variable names and APIs to fix the bug. 
However, the source statement $s$ can be an effective reference for generating correct repair templates only when $s$ \textit{shares the same syntactic structures} with the expected correct translation. If there are any structural differences between them, the solution fails to produce a valid repair patch.
For instance, as shown in Figure~\ref{Model-Example3}, BatFix identifies a suspicious statement in the C++ translation $t$: \code{int L[n][n];}, which declares a two-dimensional array but fails to initialize all its elements to \code{0} as the corresponding Python statement $s$: \code{L=[[0 for x in range(n)] for y in range(n)]} does. To fix this semantic error, BatFix first locates the corresponding Python statement $s$ aligned with $t$ (①), and then designs the following patch generation template by referencing $s$ (②):

\begin{itemize}
  \item \codeWoQuot{int L=[[0 for variable\#0 in function\#0] for variable\#1 in function\#1];}
\end{itemize}

However, this template, derived from the source Python code, does not conform to C++ grammar and significantly deviates from the correct translation. Consequently, after filling variable and function names into this flawed template (③), all the repair patches based on this template are predictably incorrect (④).

\subsection{Introducing Translation Rules to Address Limitations in Current Methods}
\label{subsec:idea}

To address the limitations of existing methods, we propose leveraging code translation rules as references for building code alignments and designing repair patch templates.

Code translation rules define the code transformation relationships between different PLs while preserving equivalent functionality. These rules can serve as references for both code alignment and translation error repair.
Specifically, the relationships defined by these rules can help identify source code and its functionally equivalent translations, as equivalent translations can satisfy these relationships, whereas non-equivalent ones fail.
Moreover, given the source code, translation rules can provide insights into how to fix its erroneous translations by inferring the code structure of its correct translation.

Adopting translation rules as references for building code alignment has great potential to address the limitations of existing CFG-based and one-shot-prompting methods in building code alignment for locating semantic errors.
Firstly, translation rules can capture mappings between significantly divergent code structures across different PLs.
Therefore, referencing these translation rules can help identify the alignments between translation pairs with different CFG structures, which CFG-based BatFix struggles to handle.
Secondly, unlike the one-shot example used by TransMap as the reference for code alignment, translation rules can provide more tailored references for aligning each specific translation pair. Specifically, given a translation pair for alignment, we can select the translation rules suitable for the source code to serve as references for building alignments between them. In contrast to TransMap's single prompt example, such a dynamic assignment strategy offers more context-aware guidance for building code alignment.

In addition, translation rules can assist in deriving patch generation templates, addressing the limitation of BatFix in generating repair patches for fixing semantic errors. BatFix directly references the source code structures to design patch generation templates, which makes it challenging to generate effective patches when the source code and the expected correct translation have different syntactic structures. 
In contrast, translation rules go beyond simply referencing the source code structure. They enable the derivation of the expected correct translation's structure based on the source code. The inferred translation structures can help to design patch generation templates that better align with the syntactic requirements of the target PL, ultimately facilitating the creation of more effective repair patches.

In summary, leveraging code translation rules as references offers an effective solution to address the challenges faced by BatFix and TransMap in locating and fixing translation semantic errors.

\section{Methodology}
\label{sec:methodology}

\begin{figure*}[t]
\centering
\includegraphics[width=1.0\linewidth]{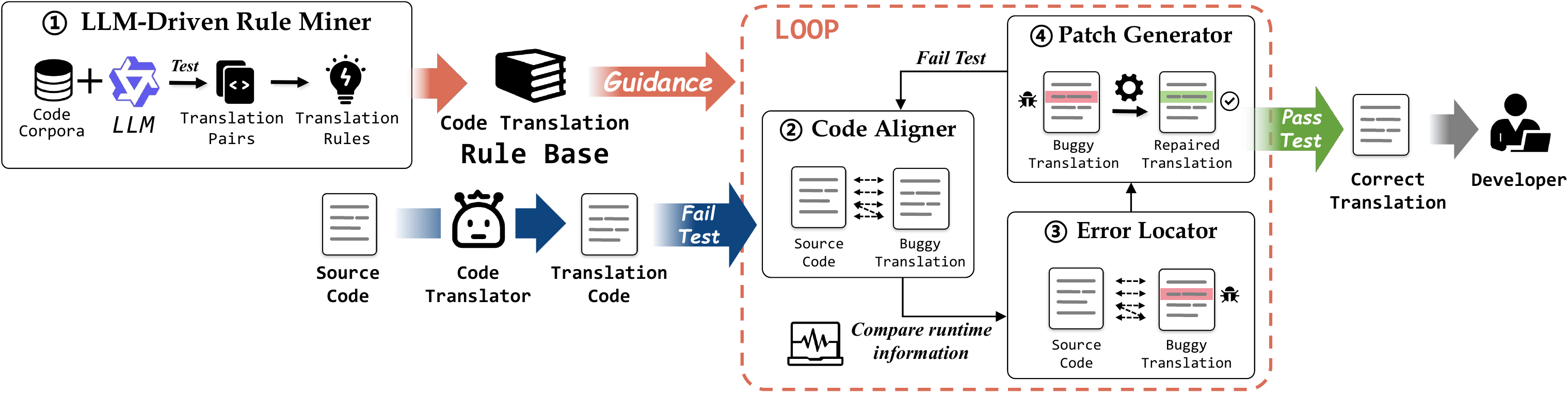}
\caption{Overview of RulER.}
\label{overview}
\end{figure*}

In this work, we proposed an automated Rule-based semantic Error localization and Repair method for code translation (RulER), which provides an end-to-end solution for debugging code translations.
Figure~\ref{overview} shows the overview of RuLER, which is composed of four modules working: \textit{LLM-Driven Rule Miner}, \textit{Code Aligner}, \textit{Error Locator}, and \textit{Patch Generator}. 
To mitigate the high cost of manually designing code translation rules, the \textit{LLM-Driven Rule Miner} leverages the LLM, which can generate code translations between various PLs, to assist in constructing translation rules.
Specifically, by filtering the correct translation of LLMs generated for a pre-defined diverse program corpus with test-based validation, \textit{LLM-Driven Rule Miner} can automatically extract massive code translation rules from these diverse correct translations.
Referring to these mined rules, the \textit{Code Aligner} module aligns the statements between the input source code and buggy translation if they satisfy the relationships defined by the rules. Subsequently, the \textit{Error Locator} module locates semantic errors by analyzing runtime divergences between aligned statements. Finally, the \textit{Patch Generator} module uses the translation rules to derive patch generation templates and synthesize repair patches to fix the erroneous translations. The three modules of \textit{Code Aligner}, \textit{Error Locator}, and \textit{Patch Generator} will operate iteratively in a loop until the repaired translation passes all tests successfully, which is returned to the developer. Subsequent sections provide detailed explanations of each module.

\subsection{LLM-Driven Rule Constructions}
\label{LLM-Driven-Rule-Miner}

The LLM-Driven Rule Miner module aims to construct diverse code translation rules across different PLs. Such rules will be applied by subsequent modules as guidance to build code alignments for locating bugs and derive templates to generate repair patches. To construct these rules automatically, this module collects plentiful correct translation pairs generated by LLM. Intuitively, this module tries to ``export'' LLM's knowledge on code translation into explicit translation rules. 

\subsubsection{Collecting Statement-Level Translation Pairs}

\begin{figure*}[t]
\centering
\includegraphics[width=1.0\linewidth]{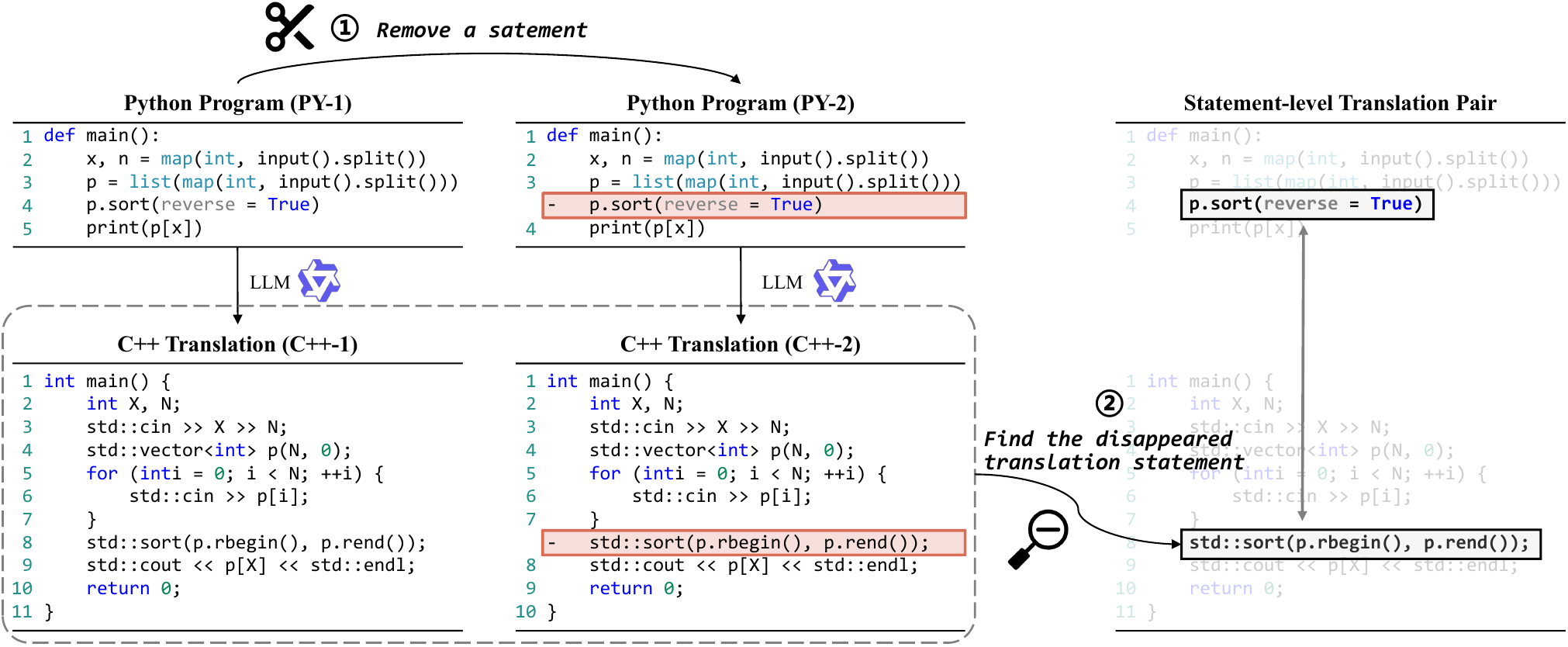}
\caption{A running example of collecting a statement-level translation pair by removing the code statement.}
\label{remove-based-example}
\end{figure*}

Statement-level translation pairs, consisting of individual statements in the source PL and corresponding translated statements in the target PL, capture the mappings between code statements with equivalent functionality across different PLs. Such pairs reflect correspondences in code structures and the APIs used within equivalent code statements, which are insightful for designing code translation rules. LLM-Driven Rule Miner aims to collect a large number of verified statement-level translation pairs, providing the foundation for constructing translation rules.

To achieve this, we propose a removal-based approach to extract statement-level translation pairs from correctly translated code snippets generated by LLMs. The key idea is to remove a specific statement from the source code and then identify the corresponding translated statements that disappear in the translations generated by LLM. By pairing the removed statement and its translated counterpart, we collect a statement-level translation pair, where the paired statements are expected to share equivalent functionality. 
This idea is inspired by the prediction interpretation method for machine-learning classifiers, which indicates the importance of each input feature to the classifier's final output by randomly masking input features~\cite{RISE, SHAP, LIME}. 

Figure~\ref{remove-based-example} illustrates this idea with an example. Given a Python code (PY-1) and its C++ translation generated by LLM (C++-1), we remove a specific statement \code{p.sort(reverse=True)} from PY-1. This results in a modified version of the Python code (PY-2), along with a new C++ translation generated by LLM (C++-2). If both C++-1 and C++-2 are verified to be correct, we compare the two translations and find that the statement \code{std::sort(p.rbegin(), p.rend());} in C++-1 has disappeared in C++-2. This disappearance indicates the correspondence between the removed Python statement and the disappeared C++ statement. Thus, we obtain a statement-level translation pair of \code{p.sort(reverse=True)} in Python and \code{std::sort(p.rbegin(), p.rend());} in C++.

\begin{algorithm}[t]
\caption{Process of extracting statement-level translation pairs}
\label{alg:RemoveBasedMapping}
\small
\SetKwInOut{Data}{Data}
\SetKwInOut{Result}{Result}
\Data{$S$: the source program written in $\mathcal{L}_S$.}
\Result{$\mathbb{P}$: set of statement translation pairs from $\mathcal{L}_S$ to $\mathcal{L}_T$;}
$\mathbb{P}$ = \{ \} \\
\While{$S$ is not empty}
{   
    $s$ = \textsc{GetLastStatement}($S$) \\
    $S'$ = \textsc{Remove}($S$, $s$) \\
    $T$ = \textsc{LLM}($S$) \\
    $T'$ = \textsc{LLM}($S'$) \\
    \If{\textsc{ValidateTranslation}($S$, $T$) and \textsc{ValidateTranslation}($S'$, $T'$)}
    {
         $\mathbb{T}$ = \textsc{GetDelta}($T$, $T'$) \\
         $\mathbb{P}$.add([$s$, $\mathbb{T}$]) \\
    }
}
\Return{$\mathbb{P}$}
\end{algorithm}

Algorithm~\ref{alg:RemoveBasedMapping} illustrates how RulER extracts statement-level translation pairs. Specifically, given a source program $S$ in the source PL $\mathcal{L}_S$, we get the last statement of $S$, denoted as $s$ (Line 3), and then remove it from $S$, resulting in a modified program $S'$ (Line 4)\footnote{We remove the statements in $S$ from last to first to avoid breaking the data dependencies in the remaining code after each removal, which may prevent it from compiling or running successfully.}. Then, we prompt an LLM to translate the original program $S$ and its modified version $S'$ into the target PL $\mathcal{L}_T$ as $T$ and $T'$ (Lines 5-6). To validate the correctness of both translations, we follow prior studies~\cite{Code-Translation-Survey1, Code-Translation-Survey2, Code-Translation-Survey3} to run $S$ ($S'$) and $T$ ($T'$) with a set of identical test inputs and check if their outputs are consistent (Line 7). If both translations are verified to be correct, we record the statements present in T but absent in $T'$ as $\mathbb{T}$ (Line 8). The pair of $s$ and the corresponding translated statement(s) in $\mathbb{T}$ is collected as a statement-level translation pair from $\mathcal{L}_S$ to $\mathcal{L}_T$ (Line 9). These translation pairs are iteratively identified and collected until all statements in the $S$ have been removed (Line 2). 

With the removal-based method, RulER can automatically collect massive verified statement-level translation pairs from LLM. These rules serve as the basis for constructing reusable code translation rules from $\mathcal{L}_S$ to $\mathcal{L}_T$.

\subsubsection{Extracting Code Translation Rules for Statement}
\label{statement-level-rule}

Based on the collected statement-level translation pairs, this step aims to extract their code structure and semantic information to construct code translation rules for statements, referred to as \textit{statement-level translation rules}.
Specifically, given a statement $s$ in source PL $\mathcal{L}_S$ and its translations $\mathbb{T}$ in target PL $\mathcal{L}_T$, we extract their Abstract Syntax Trees (ASTs) to define the code translation rule as:
\begin{equation}
\mathcal{R}_{\mathcal{L}_S \to \mathcal{L}_T}: A_{\mathcal{L}_S}\to<A^1_{\mathcal{L}_T}, A^2_{\mathcal{L}_T}, ..., A^{|\mathbb{T}|}_{\mathcal{L}_T}>
\end{equation}
where $A_{\mathcal{L}_S}$ refers to the AST of s, and $<A^1_{\mathcal{L}_T}, A^2_{\mathcal{L}_T}, ..., A^{|\mathbb{T}|}_{\mathcal{L}_T}>$ denotes the ASTs of statement $t \in \mathbb{T}$. 

We expect that code statements satisfying the ASTs defined by $\mathcal{R}_{\mathcal{L}_S \to \mathcal{L}_T}$ will exhibit equivalent functionality. However, statements with the same AST structure may have different functionalities, such as \code{int a=std::max(b, c);} and \code{int a=std::min(b, c);} in C++, which implement entirely different functionalities. As a result, a rule mapping the ASTs of \code{a=max(b, c)} in Python and \code{int a=std::max(b, c);} in C++ could lead to incorrect alignment between \code{a=max(b, c)} in Python and \code{int a=std::min(b, c);} in C++. To address this limitation, we further augment $\mathcal{R}_{\mathcal{L}_S \to \mathcal{L}_T}$ with semantic information. Specifically, if an AST node within $\mathcal{R}_{\mathcal{L}_S \to \mathcal{L}_T}$ represents a predefined identifier of its PL, we mark it with the respective identifier it represents to distinguish it from other AST nodes. 
For example, in the AST of \code{int a=std::max(b, c);}, the node representing the identifier of \code{max} is marked as \code{identifier-max} to differentiate it from other \code{identifier} nodes.
This additional semantic marking enables $\mathcal{R}_{\mathcal{L}_S \to \mathcal{L}_T}$ to more effectively differentiate between codes with identical AST structures but different semantics.

While RulER can construct lots of statement-level rules, these rules are limited to handling statements with fixed AST structures and cannot be applied to statements with diverse expression variations and tokens. For instance, given the translation rule \textit{Rule-x} mapping \code{t=N} in Python with \code{t=N;} in C++, RulER cannot apply it to match \code{t=N+1} in Python with \code{t=N+1;} in C++.

\subsubsection{Extracting Code Translation Rules for Expression and Token}

To address the limitation of statement-level rules mentioned in Section~\ref{statement-level-rule}, RulER further mines code translation rules for expressions and tokens, referred to as \textit{expression- and token-level translation rules}.
Such rules help extend existing statement-level rules to handle a broader range of code statements involving different expressions or tokens. 
For instance, if an expression-level translation rule, \textit{Rule-y}, is constructed to map the expression \code{N+1} in Python with \code{N+1} in C++, we can combine it with \textit{Rule-x}, which maps \code{t=N} in Python with \code{t=N;} in C++ to synthesize a new \textit{Rule-z} capable of matching \code{t=N+1} in Python with \code{t=N+1;} in C++. The steps for synthesizing rules will be introduced in Section~\ref{Code-Alignment}.

\begin{figure*}
\centering
\includegraphics[width=1.0\linewidth]{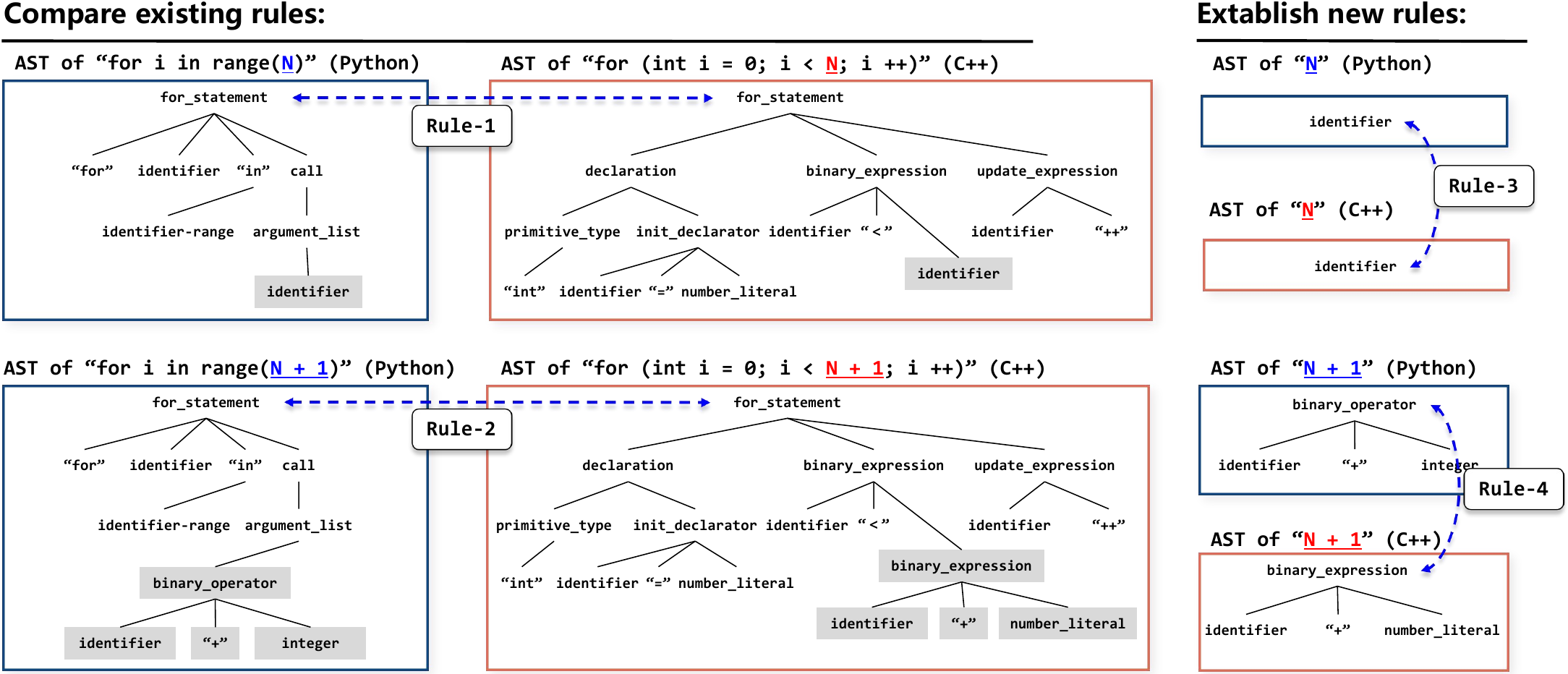}
\caption{A running example of building expression- and token-level rules by comparing statement-level rules.}
\label{ExpressionRule}
\end{figure*}

As RulER has mined large-scale statement-level rules, expression- and token-level rules can be easily derived by comparing existing statement-level rules. Specifically, given two statement-level rules, \textit{Rule-A} and \textit{Rule-B}, if their respective $A_{\mathcal{L}_S}$ share the same root node and differ by only a single subtree, an expression- or token-level code translation rule can be constructed as follows:
\begin{equation}
\mathcal{R}_{\mathcal{L}_S \to \mathcal{L}_T}: A^{sub}_{\mathcal{L}_S}\to <A^{sub^1}_{\mathcal{L}_T}, A^{sub^2}_{\mathcal{L}_T}, ..., A^{sub^x}_{\mathcal{L}_T}>
\end{equation}
where, $A^{sub}_{\mathcal{L}_S}$ is the different subtree in $A_{\mathcal{L}_S}$ between \textit{Rule-A} and \textit{Rule-B}, and $<A^{sub^1}_{\mathcal{L}_T}, A^{sub^2}_{\mathcal{L}_T}, ..., A^{sub^x}_{\mathcal{L}_T}>$ represents the different subtrees in $<A^1_{\mathcal{L}_T}, A^2_{\mathcal{L}_T}, ..., A^{|\mathbb{T}|}_{\mathcal{L}_T}>$ between \textit{Rule-A} and \textit{Rule-B}.

Figure~\ref{ExpressionRule} provides a running example for extracting expression- and token-level code translation rules. Given two statement-level rules: 

\begin{itemize}
  \item \textit{Rule-1} maps \code{for i in range(N)} to \code{for(int i=0; i<N; i++)} from Python to C++.
  \item \textit{Rule-2} maps \code{for i in range(N+1)} to \code{for(int i=0; i<N+1; i++)} from Python to C++.
\end{itemize}

The difference between \textit{Rule-1} and \textit{Rule-2} lies in their loop termination condition: \code{N} versus \code{N+1}. By comparing the two Python ASTs of \code{for i in range(N)} and  \code{for i in range(N+1)}, RulER identifies the different subtrees (highlighted in gray) corresponding to termination conditions \code{N} and \code{N+1} (highlighted in blue). Similarly, comparing the two C++ ASTs also reveals the different subtrees (highlighted in gray) corresponding to the termination conditions (highlighted in red). According to the different subtrees, RulER extracts two token- and expression-level translation rules as: 

\begin{itemize}
  \item \textit{Rule-3} maps \code{N} to \code{N} from Python to C++.
  \item \textit{Rule-4} maps \code{N+1} to \code{N+1} from Python to C++.
\end{itemize}

Finally, all the constructed statement-level, expression-level, and token-level translation rules are integrated into a rule base, serving as the basis for subsequent \textit{Code Aligner} and \textit{Patch Generator} modules to establish code alignments and derive patch generation templates.

\subsection{Code Alignment Based on Mined Rules}

The Rule Miner module applies the code translation rules produced by \textit{Rule Miner} to align statements within the source program and target translation. The process involves searching for applicable code translation rules and aligning statements that satisfy these rules.

\subsubsection{Searching for Applicable Code Translation Rules}
\label{Code-Alignment}

Given a statement $s$ in the source PL $\mathcal{L}_S$, if there is a translation rule whose $A_{\mathcal{L}_S}$ strictly matches the AST structure of $s$, this rule is considered applicable to $s$. If there is no applicable rule for $s$ in the rule base, RulER activates the \textit{rule synthesis mechanism}, which applies existing translation rules to synthesize new rules tailored to $s$.

\begin{figure*}
\centering
\includegraphics[width=1.0\linewidth]{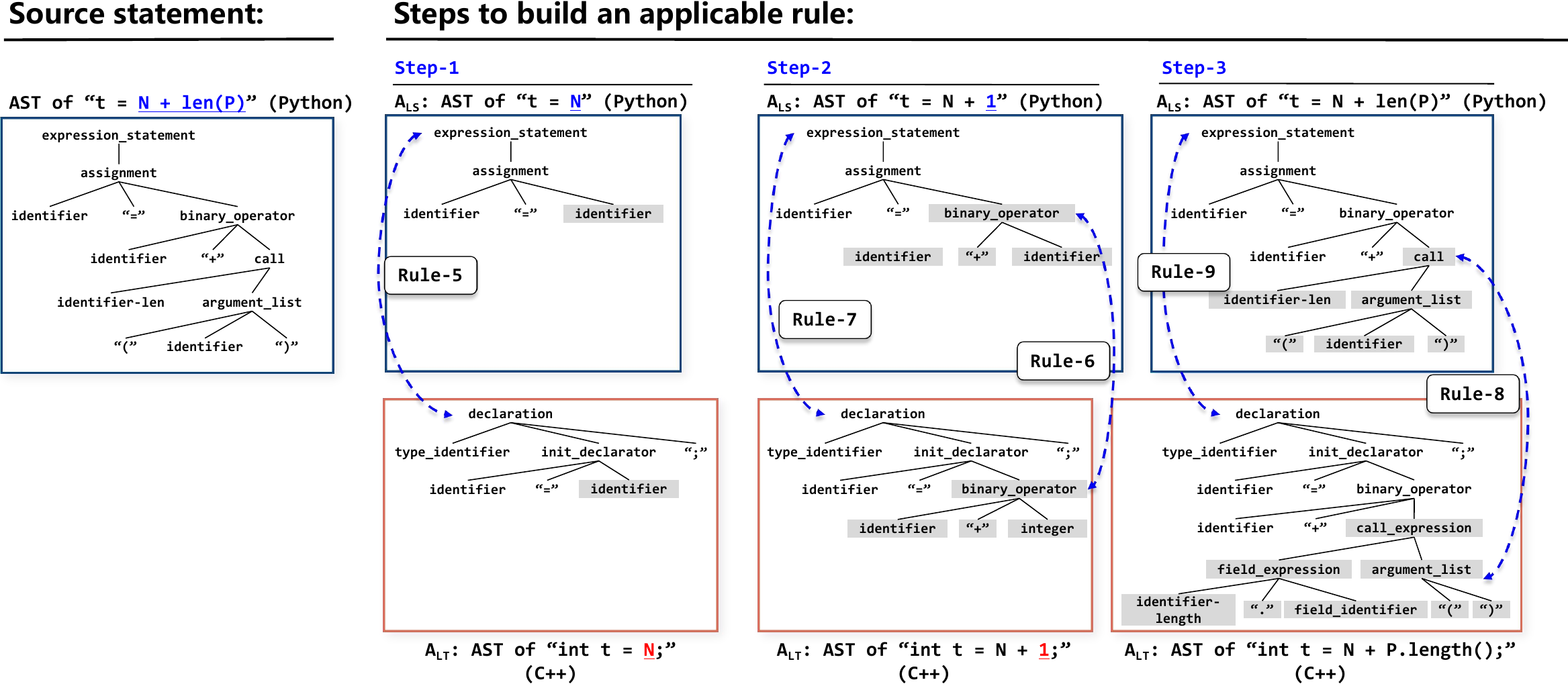}
\caption{A running example of constructing a new translation rule by combining existing rules when there is no applicable rule for the source statement.}
\label{BuildNewRule}
\end{figure*}

Figure~\ref{BuildNewRule} shows a running example of the rule synthesis mechanism to construct a new translation rule. For the source statement \code{t=N+len(P)} in Python, which lacks an applicable rule from Python to C++, RulER starts by finding \textit{Rule-5}, which maps \code{t=N} in Python to \code{t=N;} in C++, because the $A_{\mathcal{L}_S}$ of Rule-5 shares the same root node as the AST of \code{t=N+len(P)}. 
RulER then performs the following operations in sequence:

\begin{enumerate}
  \item \textbf{Compare} the AST of \code{t=N+len(P)} with $A_{\mathcal{L}_S}$ of \textit{Rule-5} to find the different subtree in the AST of \code{t=N+len(P)}, which is rooted at the \code{binary\_operator} node.
  \item \textbf{Search} for an expression-level rule, \textit{Rule-6}, whose $A_{\mathcal{L}_S}$ has a root node of \code{binary\_operator}.
  \item \textbf{Replace} the different subtrees in \textit{Rule-5} with the ASTs of \textit{Rule-6} to synthesizes \textit{Rule-7}, which maps \code{t=N+1} in Python to \code{t=N+1;} in C++.
\end{enumerate}

RulER iteratively repeats the above ``\textit{compare-search-replace}'' operations until it synthesizes a rule applicable to the Python statement \code{t=N+len(P)}. Ultimately, RulER synthesizes \textit{Rule-9}, which maps the source Python statement \code{t=N+len(P)} to corresponding C++ translation \code{t=N+P.length();}.

\subsubsection{Building Code Alignments}

For each statement $s$ in the source program $S$, if a set of translation rules $\mathbb{R}$ is available, RulER aligns the statement $t$ in translation $T$ with $s$ if their ASTs satisfy any rule in $\mathbb{R}$. 
Additionally, since it may be necessary to construct multiple-to-one alignments, while RulER-mined rules handle only one-to-one and one-to-multiple alignments, RulER further repeats the above alignment process for each statement in translation $T$. 

After performing the above alignment, some code may remain unaligned due to: (1) semantic errors in the translation that prevent it from matching any rule, or (2) no applicable rules. For such cases, if an unaligned source statement $s$ and an unaligned translation statement $t$ have their preceding and succeeding statements in $S$ and $T$ aligned respectively, RulER aligns $s$ and $t$ accordingly.

\subsection{Translation Error Localization Based on Mined Rules}

The \textit{Error Locator} module aims to identify the statements in the target translation $T$ that make $T$ behave differently from the source program $S$. Based on the alignments built by \textit{Code Aligner}, this module detects runtime divergences between the aligned statements in $S$ and $T$, thereby pinpointing suspicious translation statements.

Specifically, RulER considers two types of runtime divergences as indicators of semantic errors: (1) \textbf{execution trace divergence} and (2) \textbf{variable value divergence} following \citet{BatFix} and \citet{TransMap}. 
Execution trace divergence occurs when the currently executing statements in $S$ and $T$ are not aligned with each other. Variable value divergence arises when the currently executing statements are aligned with each other, but the same variable is found to hold different values on each side.

\begin{algorithm}[t]
\caption{Process of locating semantic error}
\label{alg:ErrorLocalization}
\small
\SetKwInOut{Data}{Data}
\SetKwInOut{Result}{Result}
\SetKwFunction{getRuntimeInfomation}{getRuntimeInfomation}
\SetKwFunction{align}{align}
\SetKwFunction{equal}{equal}
\SetKwFunction{nextStep}{nextStep}
\Data{$S$: the source program; $T$: the target translation; $\mathbb{A}$: alignments between statements in $S$ and $T$; $\mathcal{I}$: test input that causes $S$ and $T$ to produce inconsistent outputs.}
\Result{$s_T$: the suspicious statement in $T$;}
$runInfo_S$, $runInfo_T$ = \textsc{Debugger}($S$, $\mathcal{I}$), \textsc{Debugger}($T$, $\mathcal{I}$) \\
$s_S^{pre}$, $s_T^{pre}$ = $None$, $None$ \\
\While{$runInfo_S$ and $runInfo_t$ are not fully traversed}
{
    $s_S$, $\mathbb{V}_S$ = \textsc{NextStep}($runInfo_S$) \\
    $s_T$, $\mathbb{V}_T$ = \textsc{NextStep}($runInfo_T$) \\
    \If{ $s_S$ and $s_T$ are not aligned according to $\mathbb{A}$}
    {
        \Return{$s_T^{last}$} \\
    }
    \If{ not \textsc{Equal}($\mathbb{V}_S$, $\mathbb{V}_T$)}
    {
        \Return{$s_T$} \\
    }
    $s_S^{pre}$, $s_T^{pre}$ = $s_S$, $s_T$
}
\end{algorithm}

Algorithm~\ref{alg:ErrorLocalization} outlines the process for locating semantic errors by identifying execution trace divergence and variable value divergence. RulER begins by using debugging tools to collect runtime information of $S$ and $T$ with the same test input that causes $S$ and $T$ to produce inconsistent outputs (Line 1). 
The runtime information of a program is a sequence of steps: $<step_1, step_2, ..., step_n>$, where each $step_i$ contains the current executing statement $s_s$ ($s_t$) in $S$ ($T$) and the current values of accessible variables $\mathbb{V}_S$ ($\mathbb{V}_T$).
RulER then traverses the executing steps of both programs in parallel, iteratively (Lines 4-5). If the current statements $s_s$ in $S$ and $s_t$ in $T$ are not aligned according to the code alignments $\mathbb{A}$ (Line 6), RulER detects an execution trace divergence and identifies the preceding statement $s_t^{last}$ in $T$ as the location of the semantic error (Line 7). Conversely, if the statements $s_s$ and $s_t$ are aligned but the same variable has different values in their corresponding $\mathbb{V}_S$ and $\mathbb{V}_T$ (Line 9), indicating a variable value divergence, RulER marks the current statement $s_t$ as the location of the semantic error (Line 10).  
If no divergences are found, RulER continues traversing until a divergence is detected or execution ends (Line 3).

\subsection{Translation Patch Generation Based on Mined Rules}

Each time the \textit{Error Locator} locates a suspicious translation statement $t$, the \textit{Patch Generator} module is triggered to generate repair patches for $t$ to fix the semantic error. This process involves locating the source statement $s$ corresponding to $t$, leveraging translation rules to infer the expected translation structure for $s$, and designing the patch generation template based on this inferred structure. These templates, derived from translation rules, enable RulER to produce repair patches that better conform to the syntax of the target PL, enhancing its effectiveness in fixing errors.

\begin{figure*}[t]
\centering
\includegraphics[width=1.0\linewidth]{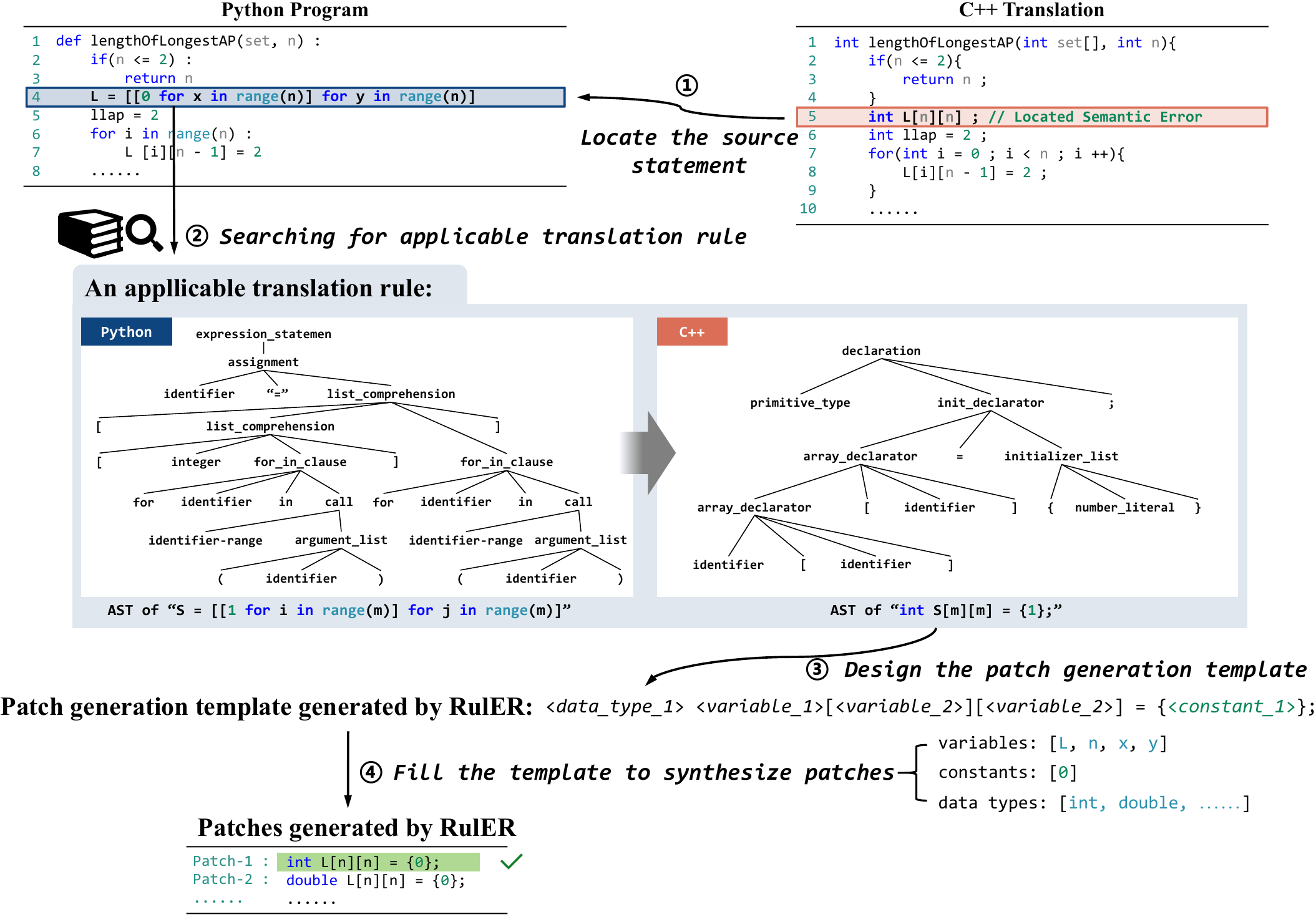}
\caption{A running example of how RulER leverages code translation rules to design the repair template and generate the repair patch for an erroneous translation.}
\label{RulER-patch}
\end{figure*}

Figure~\ref{RulER-patch} illustrates an example of how RulER synthesizes repair patches using translation rules. In this example, the \textit{Error Locator} identifies an erroneous statement $t$: \code{int L[n][n];} in the C++ translation, which fails to initialize all the elements in array \codeWoQuot{L} to \codeWoQuot{0} as the source Python program does. To repair $t$, RulER first locates the corresponding source statement $s$: \code{L=[[0 for x in range(n)] for y in range(n)]} according to code alignments(①).
Next, RulER retrieves a translation rule $\mathcal{R}$ applicable to $s$, which maps the AST of \code{S = [[1 for i in range(m)] for j in range(m)]} in Python to its equivalent C++ translation \code{int S[m][m] = \{1\};} (②). Based on the structure of \code{int S[m][m] = \{1\};}, RulER derives the patch generation template as (③):

\begin{itemize}
  \item \codeWoQuot{<data\_type\_1> <variable\_1>[<variable\_2>][<variable\_2>] = \{<constant\_1>\};}
\end{itemize}

where \codeWoQuot{<data\_type\_x>}, \codeWoQuot{<variable\_x>}, and \codeWoQuot{<constant\_x>} represent placeholders to be filled with specific data types, variable names, and constant values.
Finally, RulER fills these placeholders with elements from $s$ and $t$, i.e., \code{int}, \code{L}, \code{n}, and \code{0}, to produce a valid repair patch: 

\begin{itemize}
  \item \codeWoQuot{int L[n][n] = \{0\};}
\end{itemize}

which fixes the semantic error by initializing all elements in \codeWoQuot{L} to 0, preserving the functionality of the original Python code (④).

\section{Experimental Setup}
\label{sec:expsetup}

\subsection{Research Questions}

We comprehensively evaluate the effectiveness of RulER with four Research Questions (RQs):

\textit{\textbf{RQ1: Among the source code statements, to what extent can RulER provide applicable translation rules?}}
In this RQ, we evaluate the percentage of source code statements in the evaluation dataset for which RulER can provide applicable translation rules based on the mined rules. This evaluation reveals how extensively RulER can construct rules to handle the diverse source programs.

\textit{\textbf{RQ2: How well does RulER compare to existing methods for building code alignments?}} 
This RQ evaluates the performance of RulER's \textit{Code Aligner} module in building code alignments in terms of precision, recall, and F1 score, and compares it with two state-of-the-art code translation debugging methods, BatFix and TransMap. 

\textit{\textbf{RQ3: How well does RulER compare to existing methods for locating semantic errors?}} 
In this RQ, we evaluate the effectiveness of RulER's \textit{Error Locator} module in locating semantic errors in terms of success rate and compare it with BatFix and TransMap.

\textit{\textbf{RQ4: How well does RulER compare to BatFix and LLM for repairing semantic errors?}} 
In this RQ, we assess the success rate of RulER in repairing translations with semantic errors and compare it against BatFix. We further compare RulER with an LLM to investigate the effectiveness of repair patches generated using code translation rules mined from the LLM versus repair patches directly generated by the LLM.

\subsection{Implementation of RulER}

We implement the LLM-Driven Rule Miner and Error Locator modules in RulER as follows. 

\textbf{LLM-Driven Rule Miner module}: We use the CodeNet dataset~\cite{CodeNet}, which includes code in multiple PLs for competitive programming tasks, for rule mining. 
Specifically, we randomly select 5000 Java code snippets, 5000 Python code snippets, and 5000 C++ code snippets from CodeNet as the source code for collecting translations from Java to C++, Python to C++, and C++ to Java and Python, respectively. We choose an open-source LLM, Qwen2.5-Coder-32B-Instruct, as the code translator due to its top ranking among open-source LLMs in the EvalPlus code generation benchmark~\cite{EvalPlus}. 
As mentioned in Section~\ref{LLM-Driven-Rule-Miner}, we mine rules from correct translations generated by LLM.
To filter correct translations generated by Qwen2.5-Coder-32B-Instruct, we use all the test cases provided by CodeNet to compare whether the outputs produced by the source code and the generated translation are consistent. 
Ultimately, the LLM-Driven Rule Miner mined a substantial number of code translation rules for Java-to-C++, Python-to-C++, C++-to-Java, and C++-to-Python, as presented in Table~\ref{code-translation-rules-number}. Moreover, we evaluate the quality of these mined rules by randomly sampling 100 rules from each language pair and manually labeling their correctness. As shown in Table~\ref{code-translation-rules-accuracy}, the average accuracy of the rules mined by RulER reaches 96.5\%, demonstrating the effectiveness of RulER in extracting correct translation rules.

\begin{table}[t]
\centering\footnotesize
    \caption{Statistics of code translation rules mined by RulER.}
    \label{code-translation-rules-number}
    \begin{tabular}{ccccc}
        \toprule
        \makecell[c]{Code Translation Rules} & \makecell[c]{\# Java$\rightarrow$C++}  & \makecell[c]{\# Python$\rightarrow$C++} & \makecell[c]{\# C++$\rightarrow$Java} & \makecell[c]{\# C++$\rightarrow$Python}\\
        \midrule
        Statement-level & 5567 & 4304 & 4757 & 3272 \\
        Expression- and token-level & 4204 & 2590 & 3544 & 1944 \\
        \midrule
        Total & 9771 & 6894 & 8301 & 5216 \\
        \bottomrule
    \end{tabular}
\end{table}

\begin{table}[t]
\centering\footnotesize
    \caption{Accuracy of code translation rules mined by RulER.}
    \label{code-translation-rules-accuracy}
    \begin{tabular}{cccccc}
        \toprule
        & \makecell[c]{Java$\rightarrow$C++}  & \makecell[c]{Python$\rightarrow$C++} & \makecell[c]{C++$\rightarrow$Java} & \makecell[c]{C++$\rightarrow$Python} & Average\\
        \midrule
        Accuracy & 97.0\% & 97.0\% & 98.0\% & 94.0\% & 96.5\%\\
        \bottomrule
    \end{tabular}
\end{table}

\textbf{Error Locator module}: We leverage debugging tools specific to each PL to collect execution information of the programs. Specifically, we employ GDB\footnote{\url{https://sourceware.org/gdb/}} for C++ programs, JDB\footnote{\url{https://docs.oracle.com/javase/8/docs/technotes/tools/unix/jdb.html}} for Java programs, and PDB\footnote{\url{https://docs.python.org/3/library/pdb.html}} for Python programs. 

\subsection{Data Preparation and Annotation}
\label{dataprep}

We reuse the evaluation dataset constructed by \citet{BatFix} for evaluating BatFix.
The dataset includes 698 Java programs and 698 Python programs, along with corresponding C++ translations generated by three code translation models. The original Java and Python programs, which implement well-known algorithms, are collected by \citet{TransCoder} from GeeksforGeeks~\cite{GeeksforGeeks}. The C++ translations are generated by two code translation models, namely Transcoder~\cite{TransCoder}, Transcoder-ST~\cite{TransCoder-ST}, and a closed-source LLM, Codex~\cite{Codex}.
Given the fast-developing performance of recently developed LLMs, we further collect translations generated by Qwen2.5-Coder-32B-Instruct, the highest-ranked open-source LLM on the EvalPlus leaderboard~\cite{EvalPlus}, to enable a more comprehensive evaluation.

\begin{table}[t]
\centering\footnotesize
    \caption{Statistics of the translations with semantic errors used for evaluation.}
    \label{evaluation-dataset-number}
    \begin{tabular}{ccc}
        \toprule
        Code Translator & \# Java$\rightarrow$C++ & \# Python$\rightarrow$C++ \\
        \midrule
        TransCoder & 24 & 131 \\
        TransCoderST & 30 & 197 \\
        Codex & 20 & 44 \\
        Qwen2.5-Coder-32B-Instruct & 16 & 91 \\
        \midrule
        Total & 90 & 463 \\
        \bottomrule
    \end{tabular}
\end{table}

To identify whether the translations contain semantic errors, we executed the source and translation programs with the same test cases, comparing whether their output results are consistent.
For each source program, \citet{TransCoder} have created 10 unit test cases and manually checked their adequacy to validate translation correctness.
Inputting these 10 test cases, translations that produce outputs inconsistent with the source code for any test case are considered to contain semantic errors. Additionally, we conduct manual inspections to exclude cases where output discrepancies are caused by errors in the source code, such as integer overflow.
Finally, as shown in Table~\ref{evaluation-dataset-number}, we identify a total of 553 code translations with semantic errors for evaluation, which include 90 cases from Java to C++ and 463 cases from Python to C++.

We carefully annotated labels for the dataset to ensure a reliable evaluation. 
Two graduate students proficient in Java, Python, and C++ were employed to independently annotate the dataset. Specifically, for each pair of source and translation programs, they are required to annotate:

\begin{itemize}
  \item \textbf{Code Alignment Relationships}: For each statement in the source code, identify its corresponding translation statement(s) in the target translation. Then, annotate the alignments between their respective lines of code.
  \item \textbf{Locations of Semantic Errors}: Identify the buggy statements in translation that cause discrepancies between the outputs of the translation program and the source program, and annotate their corresponding lines of code. During this process, annotators are provided with runtime information of the source and translated programs, when they generate different outputs for identical test input, to help them identify locations where discrepancies occurred during program execution.
\end{itemize}

To evaluate the reliability of annotations, we calculate Cohen's Kappa coefficient~\cite{cohen1960coefficient}. Two annotators achieve a Cohen's Kappa value of 0.839 for code alignment annotations and 0.897 for semantic error localization annotations, demonstrating substantial agreement. Ultimately, any discrepancies between the annotations were resolved through discussion to reach a consensus.

\subsection{Methods for Comparison}
\label{sec:baselines}

We compare RulER with two state-of-the-art baseline methods, BatFix and TransMap. For localizing semantic errors in code translations, we compare RulER with both BatFix and TransMap. Since TransMap does not support automated repair, we compare RulER with BatFix and a general LLM-based patch generation method for repairing the semantic errors in code translations. The implementation details of these baselines are as follows:

\textbf{BatFix}: We adopt BatFix's replication code\footnote{\url{https://github.com/danieltrt/BatFix}} to run BatFix on our evaluation dataset, collecting its code alignment and error localization results for comparison. As for the task of repairing semantic errors, we directly use the repair patches released in BatFix's replication package, as we fail to replicate the repair functionality of BatFix based on its released code.

\textbf{TransMap}: Since the released replication of TransMap\footnote{\url{https://github.com/HALOCORE/TransMap}} is only applicable for Python to JavaScript translations, we modify its implementation to accommodate our evaluation datasets for Java-to-C++ and Python-to-C++ translations. We run TransMap with gpt-3.5-turbo to collect its code alignment and error localization results on the evaluation dataset for comparison.

\textbf{LLM-based Patch Generation}: With the advancement of LLMs, they have shown impressive performance in program repair~\cite{Automated-Program-Repair-via-Conversation}. Recent LLM-based code translation approaches also leverage LLMs to improve the quality of generated translations \cite {UniTrans, TransAGENT}. Therefore, we further compare RulER with the LLM for fixing translation errors. Specifically, following prior works on LLM-based program repair and code translation~\cite{Automated-Program-Repair-via-Conversation, APR-prompt1, APR-prompt2, APR-LLM3, TransAGENT}, we adopt a cloze-style prompt to query the LLM for generating repair patches for the suspicious lines in the translation. We select Qwen2.5-Coder-32B-Instruct for comparison, considering its notable performance on the EvalPlus leaderboard~\cite{EvalPlus}. To increase the diversity of generated patches, we set the temperature parameter to 0.8 and query the LLM 10 times for each suspicious code line in the translation following \cite{UniTrans}.

\subsection{Evaluation Metrics}

To comprehensively evaluate the performance of RulER in comparison with baseline methods, we employ the following metrics to measure its performance in providing available translation rules, building code alignments, locating and repairing translation semantic errors:

\begin{itemize}

\item We assess the proportion of source code statements for which RulER can provide applicable translation rules using the metric:
\begin{equation}
P_{applicable} = \frac{N_{applicable}}{N_{stmt}}
\end{equation}
where $N_{applicable}$ represents the number of the source code statements for which RulER can provide applicable translation rules, and $N_{stmt}$ represents the total number of source code statements used for evaluation. Given a source statement $s$, a translation rule is considered applicable to $s$ if its $A_{\mathcal{L}_S}$ strictly matches the AST of $s$.

\item To evaluate the effectiveness of RulER in building code alignments, we employ three metrics of \textit{precision}, \textit{recall}, and \textit{F1} score. Specifically, when a line in the source program should align with a line in the target program, a true positive is recorded if the method successfully aligns them together. Otherwise, it results in a false negative. If two lines that should not be aligned are incorrectly aligned by the method, it counts as a false positive. Collecting the sum of true positives, false positives, and false negatives as \textit{TP}, \textit{FP}, and \textit{FN}, we calculate the metrics of \textit{precision}, \textit{recall}, and \textit{F1} score.

\item To evaluate RulER's effectiveness in locating semantic errors, we measure the success rate of semantic error localization as: 
\begin{equation}
S_{locate} = \frac{N_{locate}}{N_{total}}
\end{equation}
where $N_{locate}$ refers to the count of translations whose semantic errors are successfully located by the method, and $N_{total}$ represents the number of all the translations containing semantic errors. For a target translation, the localization is considered successful if the suspicious code line reported by the method matches any manually annotated code lines responsible for semantic errors, thus increasing $N_{locate}$ by 1.

\item To evaluate RulER's performance in repairing semantic errors, we measure the proportion of successfully repaired translations as: 
\begin{equation}
S_{pass} = \frac{N_{pass}}{N_{total}}
\end{equation}
where $N_{pass}$ represents the count of translations that pass the test after repair, and $N_{total}$ indicates the number of all the translations with semantic errors. For each erroneous translation, if the repaired version produces consistent outputs with the source program across all 10 test cases created by \citet{TransCoder}, the repaired translation is considered to have passed the test, thus increasing $N_{pass}$ by 1.

\end{itemize}

\section{Results and Analysis}
\label{sec:expresult}

\subsection{\textbf{RQ1:} Among the source code statements, to what extent can RulER provide applicable translation rules?}
\label{sec:rq1}

\begin{table}[t]
\centering\footnotesize
    \caption{Performance of RulER to provide applicable rules for the source statements in evaluation dataset.}
    \label{rq1table}
    \begin{tabular}{cccc}
        \toprule
         & \makecell[c]{Java} & \makecell[c]{Python}  & \makecell[c]{Average}\\
        \midrule
        RulER & 94.0\% & 92.0\% & 92.6\% \\
        RulER w/o Rule Synthesis & 47.0\% & 34.4\% & 38.4\% \\
        \bottomrule
    \end{tabular}
\end{table}

\begin{figure}[t]
    \centering
    \includegraphics[width=0.65\textwidth]{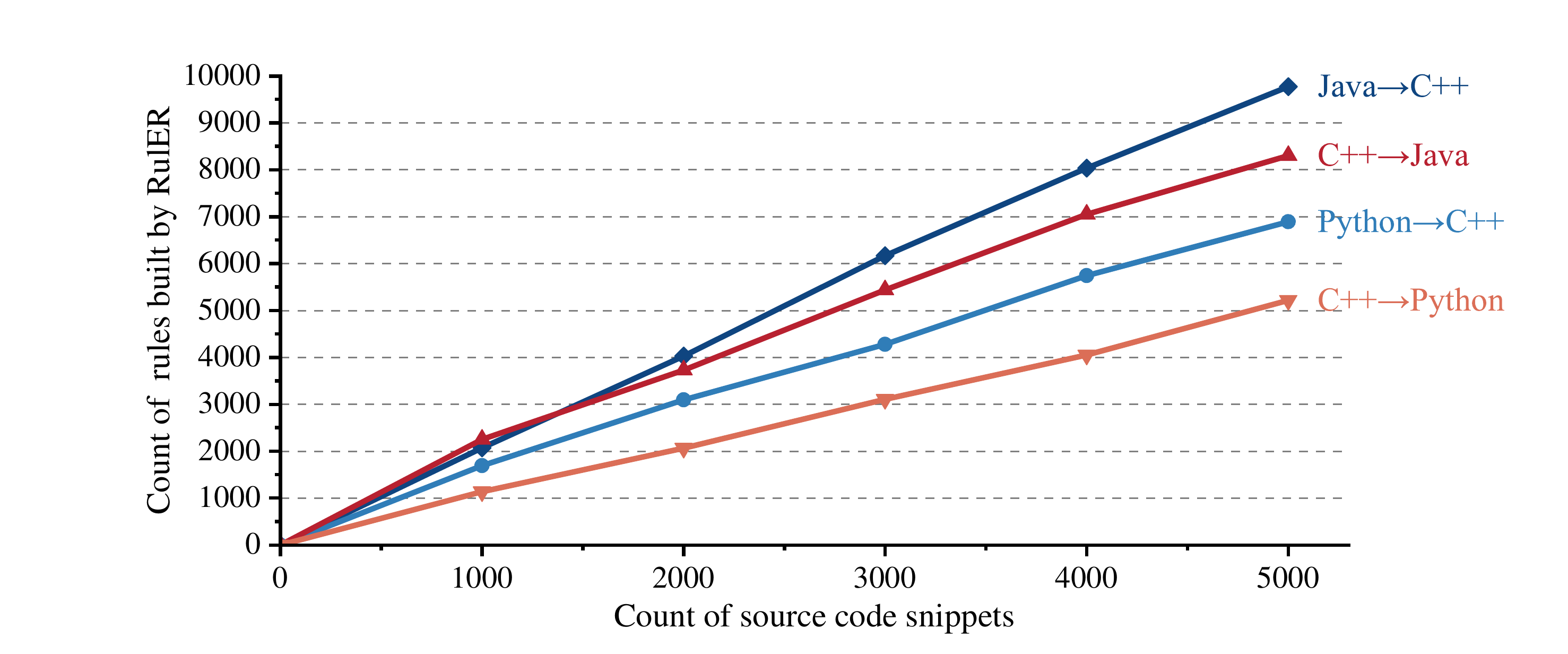}
\caption{As the number of original code snippets increases, RulER can build more code translation rules.}
\label{rq1figure}
\end{figure}

We collect all the source code statements from the evaluation dataset and filter out those with the same AST. As a result, we obtain 1,099 source code statements with unique ASTs and use RulER to generate translation rules for these statements.
Table~\ref{rq1table} presents the proportion of source code statements for which RulER can provide applicable translation rules, denoted as $P_{applicable}$. Overall, RulER provides applicable translation rules for 92.6\% of the source code statements. 
For specific PLs, RulER achieves a slightly higher $P_{applicable}$ for Java (94.0\%) compared to Python (92.0\%). This higher proportion can be attributed to the larger set of Java-to-C++ rules available compared to Python-to-C++ rules, suggesting that a richer rule base may contribute to better handling of the source code statements.

Figure~\ref{rq1figure} illustrates the number of translation rules mined by RulER when varying amounts of original code snippets are provided. We find that even though RulER has already constructed thousands of rules based on 5,000 original code snippets, it still demonstrates a consistent trend of mining more rules as more original code snippets are supplied, indicating its potential to achieve even higher $P_{applicable}$. 

\textbf{Ablation Study.} As introduced in Section~\ref{Code-Alignment}, RulER adopts the rule synthesis mechanism to synthesize new translation rules for the source code statement that lack applicable rules by combining existing ones. As shown in Table~\ref{rq1table}, the average $P_{applicable}$ achieved by RulER w/o Rule Synthesis decreases from 92.6\% to 38.4\%, highlighting the remarkable effectiveness of the rule synthesis mechanism in generating new rules beyond the rules mined by RulER.

\subsection{\textbf{RQ2:} How well does RulER compare to existing methods for building code alignments?}
\label{sec:rq2}

Table~\ref{rq2table} presents the performance of RulER and baseline methods in constructing code alignments in terms of precision, recall, and F1 score. Overall, RulER achieves F1 scores exceeding 90\% across translation pairs generated by the four translators for two PL pairs. RulER outperforms both BatFix and TransMap in F1 scores across all datasets, and achieves higher precision or recall in most cases. This demonstrates the effectiveness of the rule-based RulER in constructing more accurate code alignments across different code translators and PLs.

\begin{table*}[t]
\centering\footnotesize
    \caption{Effectiveness of different methods for building code alignments.}
    \label{rq2table}
    \begin{tabular}{cccccccc}
        \toprule
        \multirow{2}{*}{Code Translator} & \multirow{2}{*}{Method} & \multicolumn{3}{c}{Java$\rightarrow$C++} & \multicolumn{3}{c}{Python$\rightarrow$C++}\\
        \cmidrule(lr){3-5} \cmidrule(lr){6-8}
         & & \textbf{$precision$} & \textbf{$recall$} & \textbf{$F1$} & \textbf{$precision$} & \textbf{$recall$} & \textbf{$F1$} \\
        \midrule
        \multirow{3}{*}{TransCoder} & BatFix & 84.3\% & 81.0\% & 82.6\% & 84.0\% & 60.8\% & 70.5\%  \\
        & TransMap & 78.4\% & 61.3\% & 68.8\%  & 86.8\% & 78.6\% & 82.5\%  \\
        & RulER & \textbf{99.2} & \textbf{96.9\%} & \textbf{98.0\%}  & \textbf{97.9\%} & \textbf{94.3\%} & \textbf{96.1\%}  \\
        \midrule
        \multirow{3}{*}{TransCoderST} & BatFix & 82.0\% & 77.3\% & 79.6\%  & 84.7\% & 59.4\% & 69.8\%  \\
        & TransMap & 80.8\% & 56.2\% & 66.3\%  & 87.0\% & 81.5\% & 84.2\%  \\
        & RulER & \textbf{98.1\%} & \textbf{97.3\%} & \textbf{97.7\%}  & \textbf{98.0\%} & \textbf{94.2\%} & \textbf{96.1\%}  \\
        \midrule
        \multirow{3}{*}{Codex} & BatFix & 83.4\% & 79.4\% & 81.4\%  & 83.5\% & 61.9\% & 71.1\%  \\
        & TransMap & \textbf{100.0\%} & 69.3\% & 81.9\%  & 91.4\% & 86.7\% & 89.0\%  \\
        & RulER & 99.7\% & \textbf{97.6\%} & \textbf{98.6\%}  & \textbf{93.1\%} & \textbf{88.6\%} & \textbf{90.8\%}  \\
        \midrule
        \multirow{3}{*}{Qwen2.5-Coder-32B-Instruct} & BatFix & 93.4\% & 82.0\% & 87.3\%  & 87.5\% & 75.8\% & 81.2\%  \\
        & TransMap & \textbf{97.9\%} & 49.2\% & 65.5\%  & 96.2\% & \textbf{96.3\%} & 96.2\%  \\
        & RulER & 96.1\% & \textbf{90.5\%} & \textbf{93.2\%}  & \textbf{99.6\%} & 95.9\% & \textbf{97.7\%}  \\
        \bottomrule
    \end{tabular}
\end{table*}

Specifically, the CFG-based alignment method employed by BatFix achieves relatively lower F1 scores for Python-to-C++ translations compared to Java-to-C++. This is primarily because Python programs and their C++ translations typically exhibit more divergent CFG structures than those in Java and C++. In contrast, the translation rules mined by RulER effectively map between these divergent Python and C++ code structures, enabling RulER to build more accurate alignments for Python-to-C++ translations than BatFix, with the most notable improvement in F1 score reaching 26.3\% for Python-to-C++ translations generated by TransCoderST.
While TransMap demonstrates higher precision or recall than RulER for translations generated by LLMs, Codex and Qwen2.5, its alignment performance is much less effective for translations produced by non-LLM models, TransCoder and TransCoderST.
In contrast, RulER consistently achieves stable alignment performance for both LLM-generated and non-LLM-generated translations. This can be attributed to RulER's ability to provide diverse applicable rules for the source code. These rules offer more varied references for alignment than the one-shot examples used by TransMap, thereby facilitating more accurate alignments across translations generated by different models.

\begin{figure}[t]
    \centering
    \includegraphics[width=\textwidth]{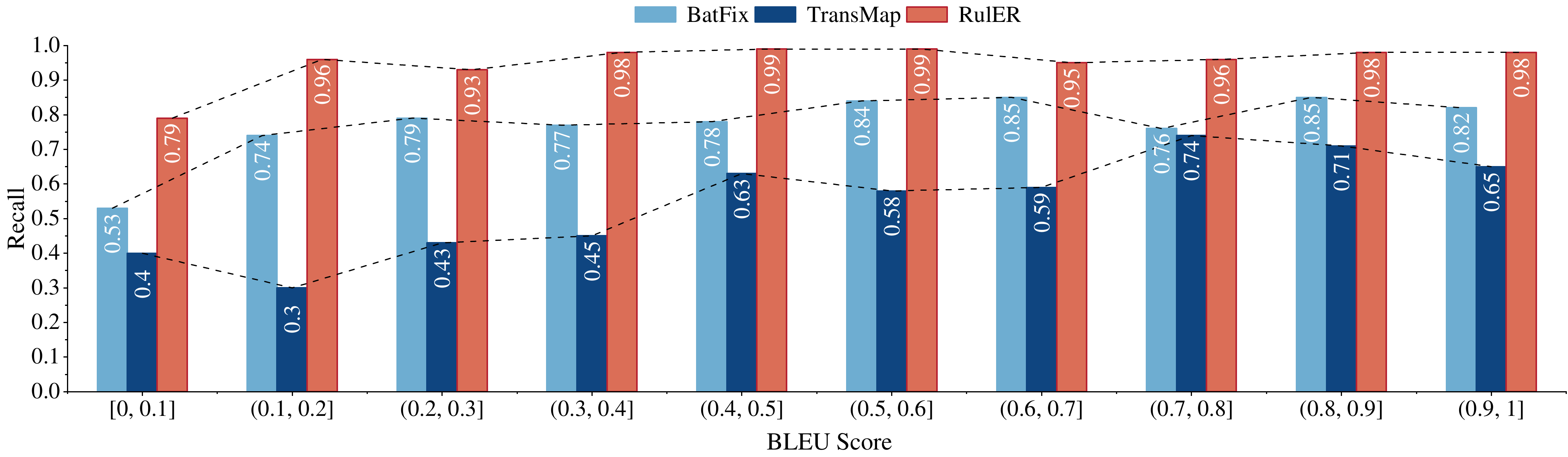}
\caption{Recall rates of different methods for building alignments between Java and C++ code with varying similarity scores.}
\label{rq2figure}
\end{figure}

To evaluate the alignment performance of each method at a finer granularity, we assessed their ability to identify alignments for source and translation code lines with different textual similarity. We hypothesize that it is more challenging to align source and translated code when they differ significantly in textual appearance. To this end, we calculate the BLEU similarity scores between manually aligned Java and C++ code lines in the evaluation dataset and categorize them into 10 classes based on their BLEU scores, ranging from 0 to 1 in increments of 0.1. For each class, we measure the recall values achieved by the methods to evaluate their ability to build alignments between such code lines.
As shown in Figure~\ref{rq2figure}, the evaluation results indicate that the recall values for all methods generally decrease as BLEU scores decline, confirming our hypothesis that textual dissimilarity poses challenges for code alignment.  Specifically, TransMap's recall drops below 0.5 when the BLEU score falls under 0.4. In contrast, BatFix maintains relatively stable recall values over 0.7 for BLEU scores above 0.1, as it relies on code structure rather than textual similarity to build code alignment. RulER consistently achieves the highest recall across all similarity levels, maintaining recall above 90\% for BLEU scores greater than 0.1. These results demonstrate that RulER outperforms BatFix and TransMap in aligning even textually dissimilar code.

\subsection{\textbf{RQ3:} How well does RulER compare to existing methods for locating semantic errors?}
\label{sec:rq3}

\begin{table*}[t]
\centering\footnotesize
    \caption{Effectiveness of different methods in locating semantic errors.}
    \label{rq3table}
    \begin{tabular}{cccccccc}
        \toprule
        \multirow{2}{*}{Code Translator} & \multirow{2}{*}{Method} & \multicolumn{3}{c}{Java$\rightarrow$C++} & \multicolumn{3}{c}{Python$\rightarrow$C++}\\
        \cmidrule(lr){3-5} \cmidrule(lr){6-8}
         & & \textbf{$N_{total}$} & \textbf{$N_{locate}$} & \textbf{\makecell[c]{$S_{locate}$}} & \textbf{$N_{total}$} & \textbf{$N_{locate}$} & \textbf{\makecell[c]{$S_{locate}$}} \\
        \midrule
        \multirow{3}{*}{TransCoder} & BatFix & 24 & 17 & 70.8\% & 131 & 53 & 40.5\% \\
        & TransMap & 24 & 16 & 66.7\% & 131 & 87 & 66.4\% \\
        & RulER & 24 & \textbf{20} & \textbf{83.3\%} & 131 & \textbf{105} & \textbf{80.2\%} \\
        \midrule
        \multirow{3}{*}{TransCoderST} & BatFix & 30 & 18 & 60.0\% & 197 & 85 & 43.1\% \\
        & TransMap & 30 & 19 & 63.3\% & 197 & 125 & 63.5\% \\
        & RulER & 30 & \textbf{22} & \textbf{73.3\%} & 197 & \textbf{157} & \textbf{79.7\%} \\
        \midrule
        \multirow{3}{*}{Codex} & BatFix & 20 & 15 & 75.0\% & 44 & 25 & 56.8\% \\
        & TransMap & 20 & 14 & 70.0\% & 44 & 32 & 72.7\% \\
        & RulER & 20 & \textbf{19} & \textbf{95.0\%} & 44 & \textbf{36} & \textbf{81.8\%} \\
        \midrule
        \multirow{3}{*}{Qwen2.5-Coder-32B-Instruct} & BatFix & 16 & 11 & 68.8\% & 91 & 36 & 39.6\% \\
        & TransMap & 16 & 9 & 56.2\% & 91 & 56 & 61.5\% \\
        & RulER & 16 & \textbf{12} & \textbf{75.0\%} & 91 & \textbf{58} & \textbf{63.7\%} \\
        \bottomrule
    \end{tabular}
\end{table*}

In this RQ, we compare the performance of RulER and baseline methods in locating semantic errors. Table~\ref{rq3table} presents the number of translation pairs whose semantic errors are correctly located by the methods and their corresponding success rates $S_{locate}$. Overall, RulER achieves $S_{locate}$ exceeding 70\% for most datasets and outperforms both BatFix and TransMap across all datasets. The higher $S_{locate}$ can be attributed to the more precise code alignments constructed by RulER. 

\begin{figure*}[t]
\centering
\includegraphics[width=0.95\linewidth]{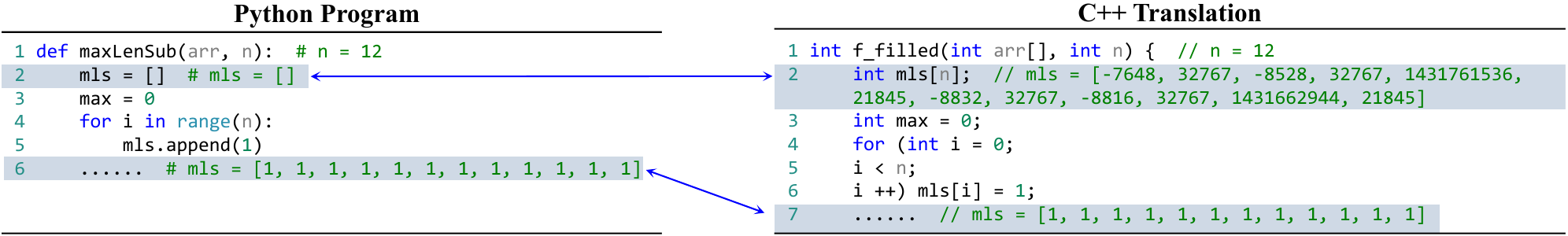}
\caption{A false localization made by RulER due to the different data types between source and target programs.}
\label{wrong-localization}
\end{figure*}

Through a manual analysis of RulER's incorrect localizations, we identified two primary reasons: (1) inaccurate code alignments and (2) the differences arising from data types between different PLs. Figure~\ref{wrong-localization} illustrates an example to show how data type differences lead to RulER's erroneous localization. In this example, the source Python code defines a list type variable \codeWoQuot{mls}, which is initialized as empty. Given that Python's list supports dynamic resizing, it then populates \codeWoQuot{mls} with \codeWoQuot{1} for \codeWoQuot{n} times through a loop. The C++ translation, however, implements \codeWoQuot{mls} as a fixed-length array of size \codeWoQuot{n}, with elements sequentially set to \codeWoQuot{1}. While functionally equivalent, the initial values of \codeWoQuot{mls} differ: one is \code{mls=[]}, while another is \code{mls=[-7648, 32767, -8528, ......]}. This value divergence causes RulER to incorrectly flag line~2 in the C++ code as containing a semantic error.

\subsection{\textbf{RQ4:} How well does RulER compare to BatFix and LLM for repairing semantic errors?}
\label{sec:rq4}

This RQ compares RulER against BatFix and an LLM in fixing semantic errors. Table~\ref{rq4table} presents the number of translations successfully repaired by each method, along with corresponding success rates $S_{pass}$. As explained in Section~\ref{sec:baselines}, we directly use the repair results released by BatFix for comparison, as we fail to reproduce its repair functionality using the provided replication code. Consequently, BatFix's results on the newly collected dataset for Qwen2.5-Coder-32B-Instruct are not available.

\begin{table*}[t]
\centering\footnotesize
    \caption{Effectiveness of different methods in fixing semantic errors.}
    \label{rq4table}
    \begin{tabular}{cccccccc}
        \toprule
        \multirow{2}{*}{Code Translator} & \multirow{2}{*}{Method} & \multicolumn{3}{c}{Java$\rightarrow$C++} & \multicolumn{3}{c}{Python$\rightarrow$C++}\\
        \cmidrule(lr){3-5} \cmidrule(lr){6-8}
         & & \textbf{$N_{total}$} & \textbf{$N_{pass}$} & \textbf{\makecell[c]{$S_{pass}$}} & \textbf{$N_{total}$} & \textbf{$N_{pass}$} & \textbf{\makecell[c]{$S_{pass}$}} \\
        \midrule
        \multirow{3}{*}{TransCoder} & BatFix & 24 & 13 & 54.2\% & 131 & 3 & 2.3\% \\
        & Qwen2.5-Coder-32B-Instruct & 24 & 7 & 29.2\% & 131 & 56 & 42.7\% \\
        & RulER & 24 & \textbf{19} & \textbf{79.2\%} & 131 & \textbf{67} & \textbf{51.1\%} \\
        \midrule
        \multirow{3}{*}{TransCoderST} & BatFix & 30 & 13 & 43.3\% & 197 & 5 & 2.5\% \\
        & Qwen2.5-Coder-32B-Instruct & 30 & 12 & 40.0\% & 197 & 86 & 43.7\% \\
        & RulER & 30 & \textbf{15} & \textbf{50.0\%} & 197 & \textbf{98} & \textbf{49.7\%} \\
        \midrule
        \multirow{3}{*}{Codex} & BatFix & 20 & 13 & 65.0\% & 44 & 11 & 25.0\% \\
        & Qwen2.5-Coder-32B-Instruct & 20 & 5 & 25.0\% & 44 & 4 & 9.1\% \\
        & RulER & 20 & \textbf{17} & \textbf{85.0\%} & 44 & \textbf{22} & \textbf{50.0\%} \\
        \midrule
        \multirow{3}{*}{Qwen2.5-Coder-32B-Instruct} & BatFix & 16 & N/A & N/A & 91 & N/A & N/A\\
        & Qwen2.5-Coder-32B-Instruct & 16 & 0 & 0.0\% & 91 & 1 & 1.1\% \\
        & RulER & 16 & \textbf{7} & \textbf{43.8\%} & 91 & \textbf{22} & \textbf{24.2\%} \\
        \bottomrule
    \end{tabular}
\end{table*}

Overall, RulER outperforms both BatFix and Qwen2.5-Coder-32B-Instruct by repairing more translations across all datasets. While BatFix shows some effectiveness in fixing translations from Java to C++, its performance drops significantly when handling Python-to-C++ translations. Particularly, BatFix achieves $S_{pass}$ less than 3\% for erroneous Python-to-C++ translations generated by TransCoder and TransCoderST. 
This underscores the limitation of BatFix in designing effective repair templates for the target PL due to its sole consideration of the grammar of the source PL, particularly when there are significant grammatical differences between the source and target PLs, such as Python and C++. To address this issue, RulER takes a further step by applying translation rules to the source Python code to infer the structures of their C++ translations. These inferred code structures for the target PL help RulER to design repair templates that better align with the target PL grammar, thereby generating more effective repair patches to fix the errors.

While Qwen2.5-Coder-32B-Instruct demonstrates some effectiveness in repairing code translations generated by TransCoder and TransCoderST, its performance is notably weaker when addressing translations generated by LLMs of Codex or even Qwen2.5-Coder-32B-Instruct itself. This discrepancy can be attributed to differences in the types of semantic errors present in the translations. Specifically, semantic errors in translations generated by TransCoder and TransCoderST often involve issues like uninitialized variables and incorrect conditional statements, which are relatively easier to identify and fix. In contrast, translations generated by Codex and Qwen2.5-Coder-32B-Instruct tend to contain more implicit, data-type-related errors, such as data overflow and mismatched return types. These issues are harder for the LLM to identify, leading it to retain the original incorrect translations without making any modifications.
In comparison, RulER leverages its extensive library of translation rules to provide multiple repair options covering various data types. This makes RulER more effective in generating patches with the correct data types to fix these data-type-related errors.

\begin{figure}[t]
    \centering
    \begin{subfigure}[b]{0.22\textwidth}
        \centering
        \includegraphics[width=\textwidth]{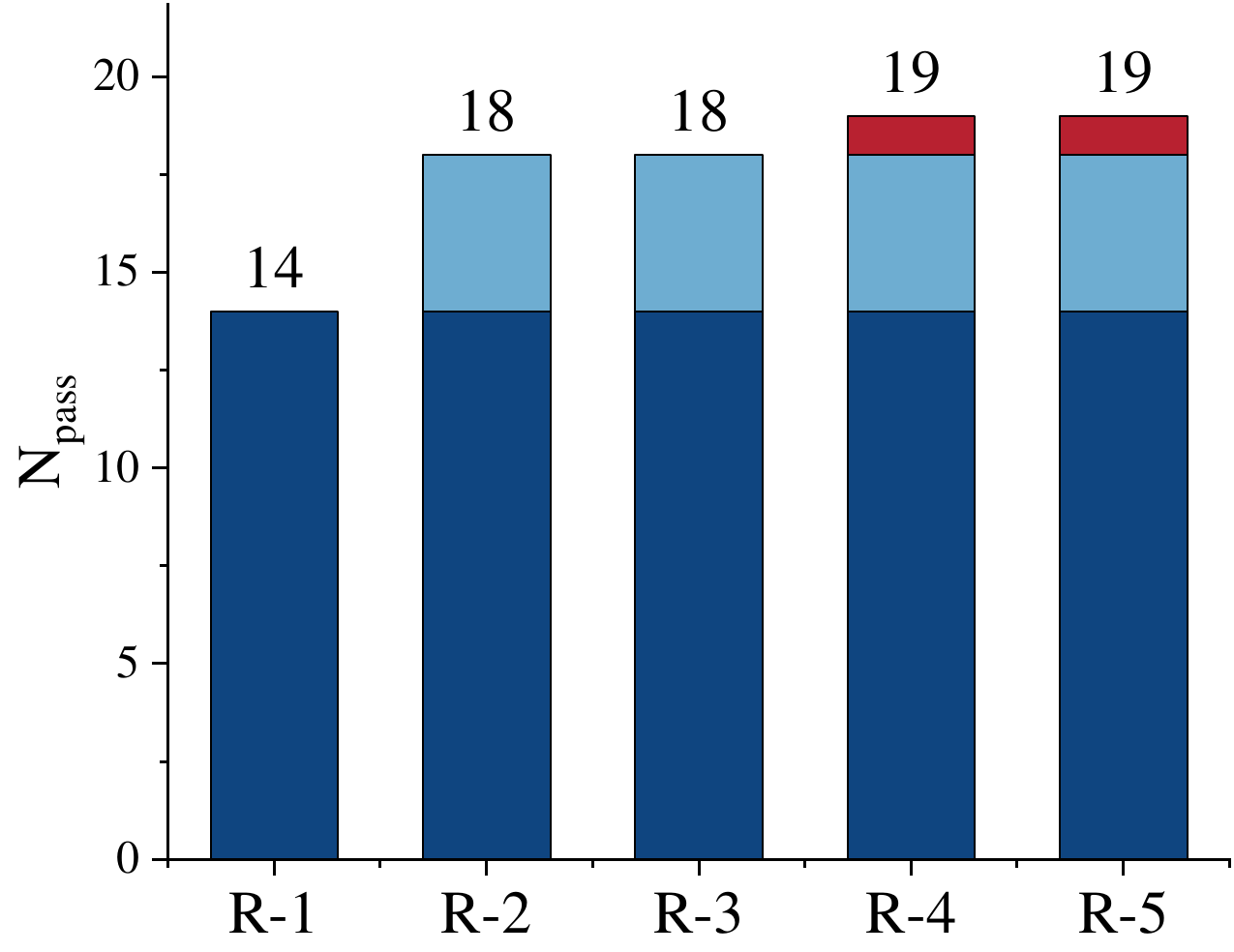}
        \caption{\tiny TransCoder (Java$\rightarrow$C++)}
        \label{rq3figure1}
    \end{subfigure}
    \hfill
    \begin{subfigure}[b]{0.22\textwidth}
        \centering
        \includegraphics[width=\textwidth]{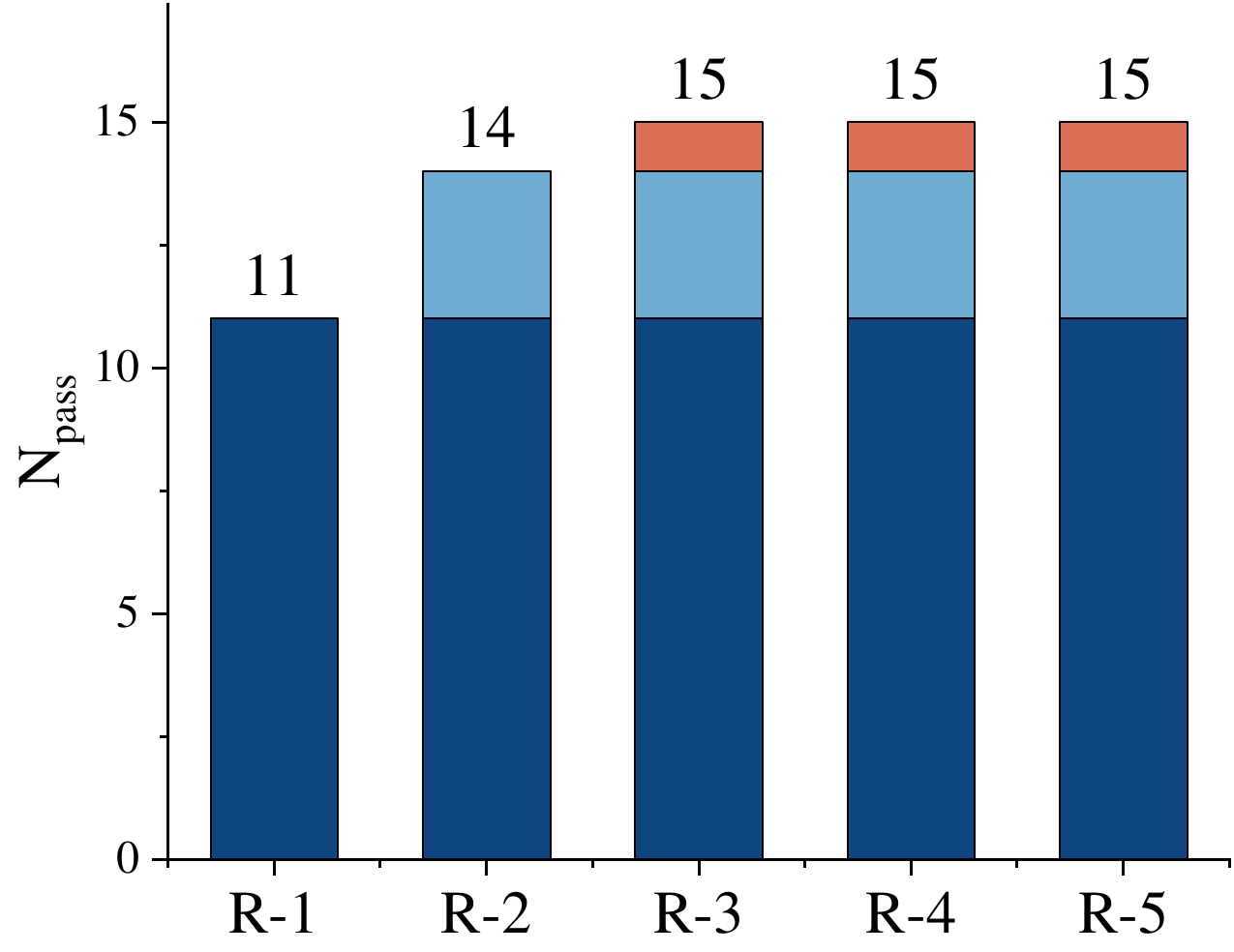}
        \caption{\tiny TransCoderST (Java$\rightarrow$C++)}
        \label{rq3figure2}
    \end{subfigure}
    \hfill
    \begin{subfigure}[b]{0.22\textwidth}
        \centering
        \includegraphics[width=\textwidth]{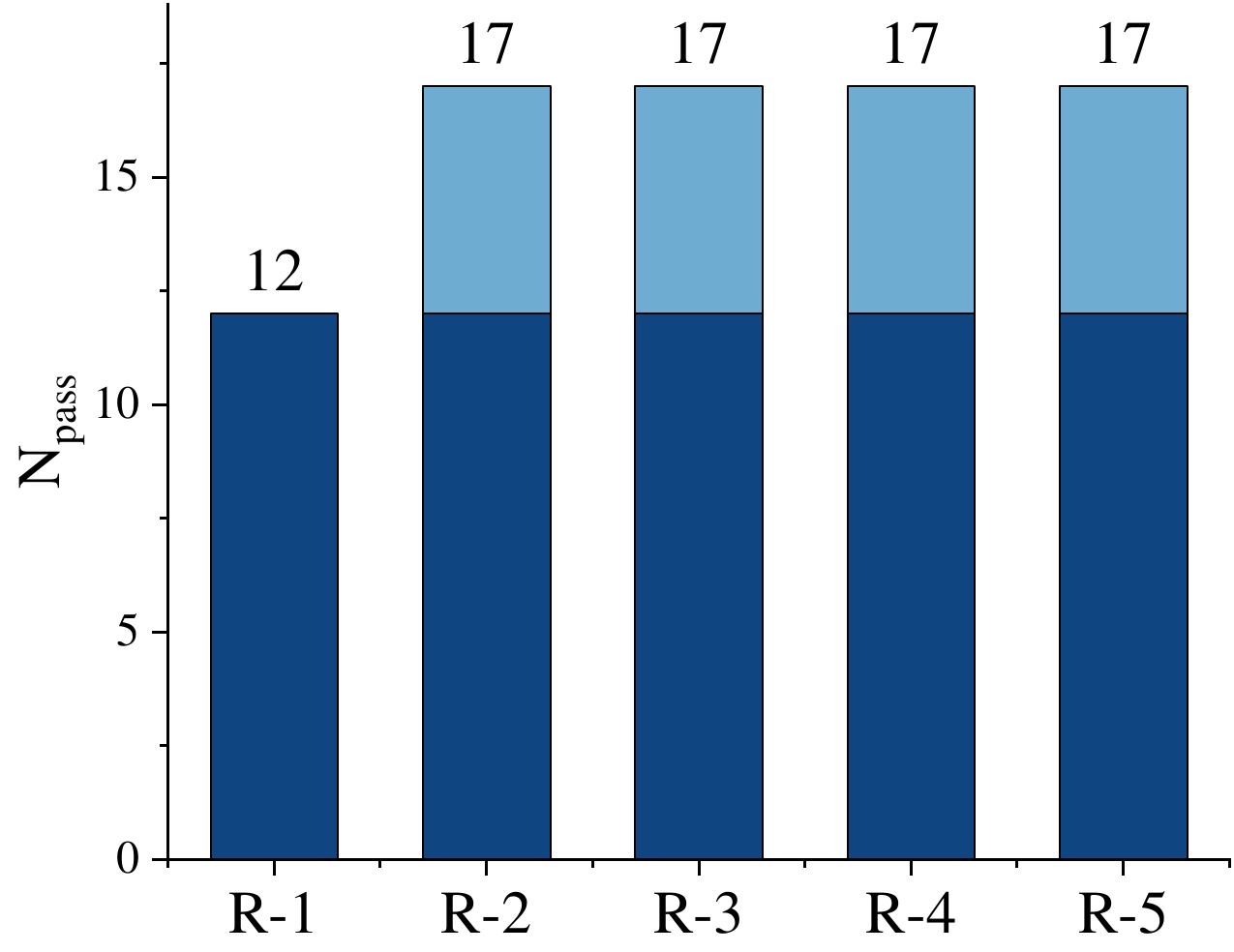}
        \caption{\tiny Codex (Java$\rightarrow$C++)}
        \label{rq3figure3}
    \end{subfigure}
    \hfill
    \begin{subfigure}[b]{0.22\textwidth}
        \centering
        \includegraphics[width=\textwidth]{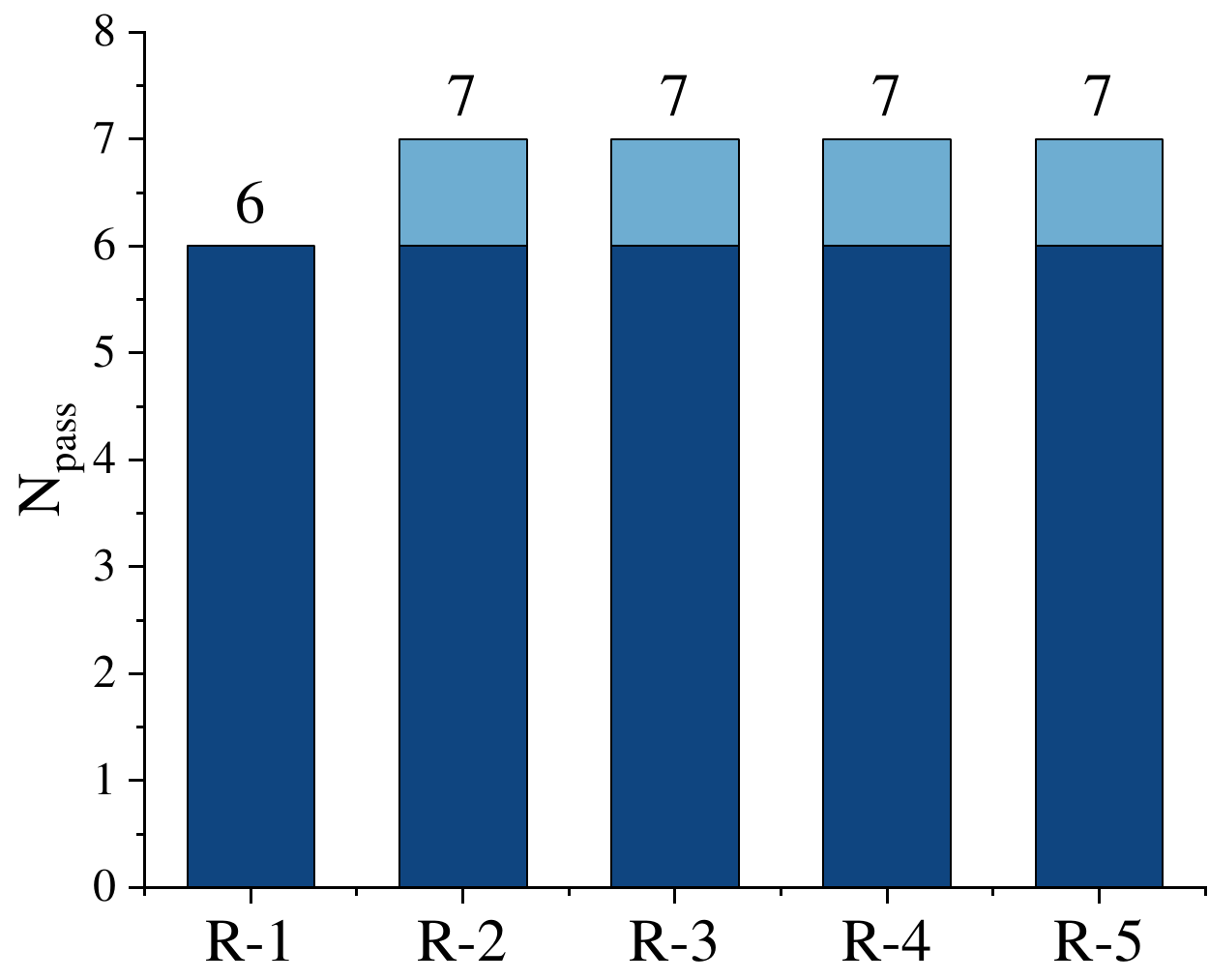}
        \caption{\tiny Qwen2.5 (Java$\rightarrow$C++)}
        \label{rq3figure4}
    \end{subfigure}
    \hfill
    \begin{subfigure}[b]{0.22\textwidth}
        \centering
        \includegraphics[width=\textwidth]{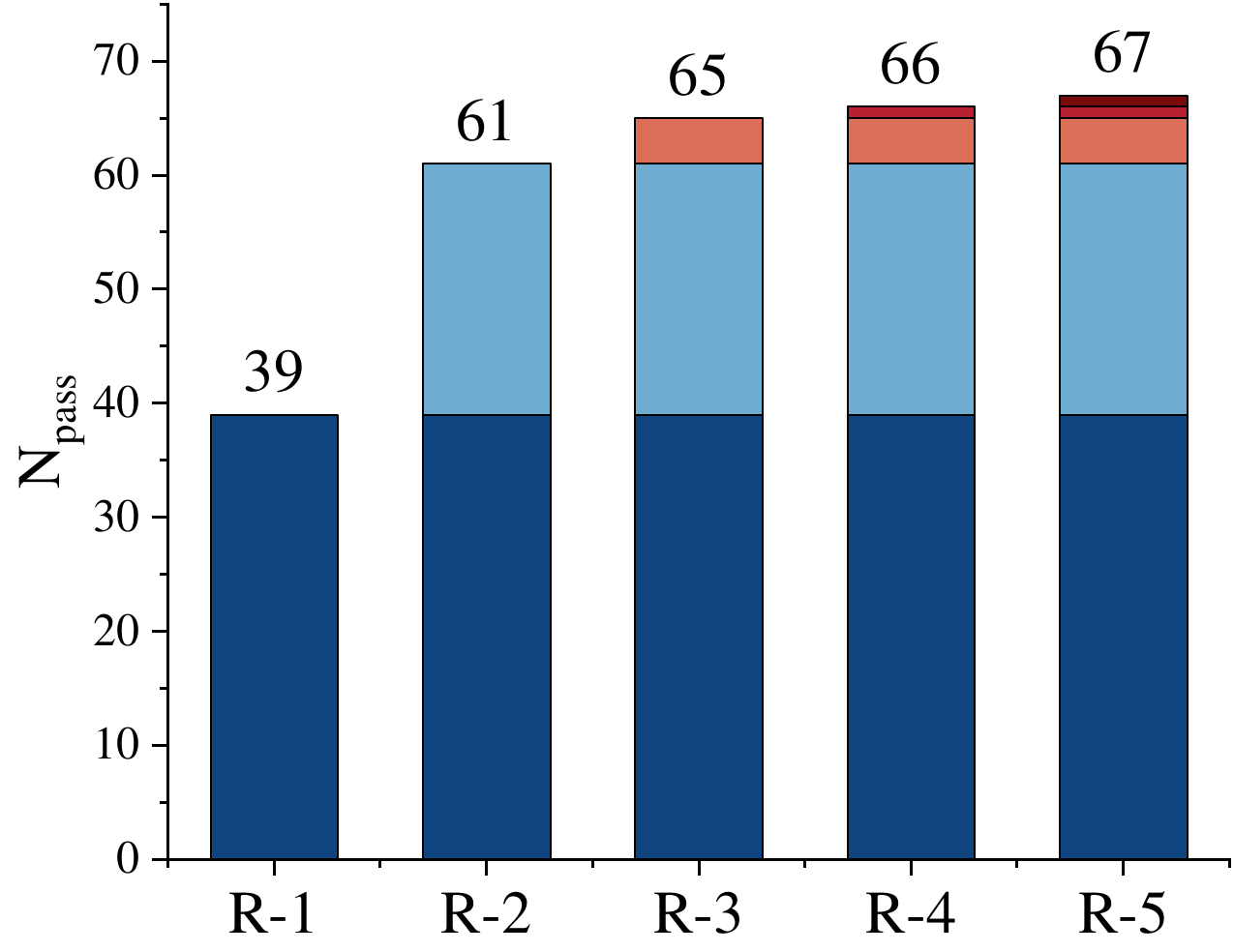}
        \caption{\tiny TransCoder (Python$\rightarrow$C++)}
        \label{rq3figure5}
    \end{subfigure}
    \hfill
    \begin{subfigure}[b]{0.22\textwidth}
        \centering
        \includegraphics[width=\textwidth]{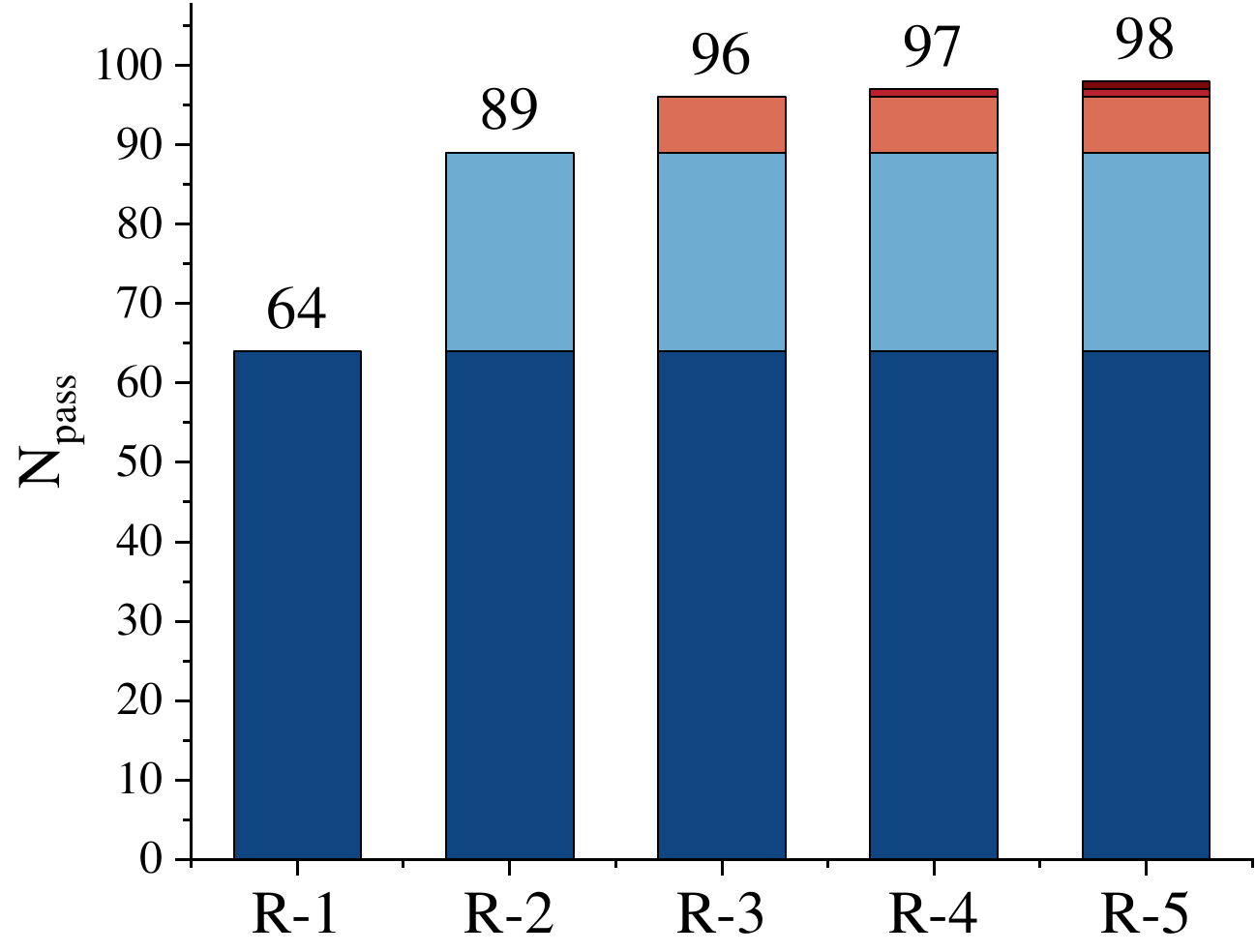}
        \caption{\tiny TransCoderST (Python$\rightarrow$C++)}
        \label{rq3figure6}
    \end{subfigure}
    \hfill
    \begin{subfigure}[b]{0.22\textwidth}
        \centering
        \includegraphics[width=\textwidth]{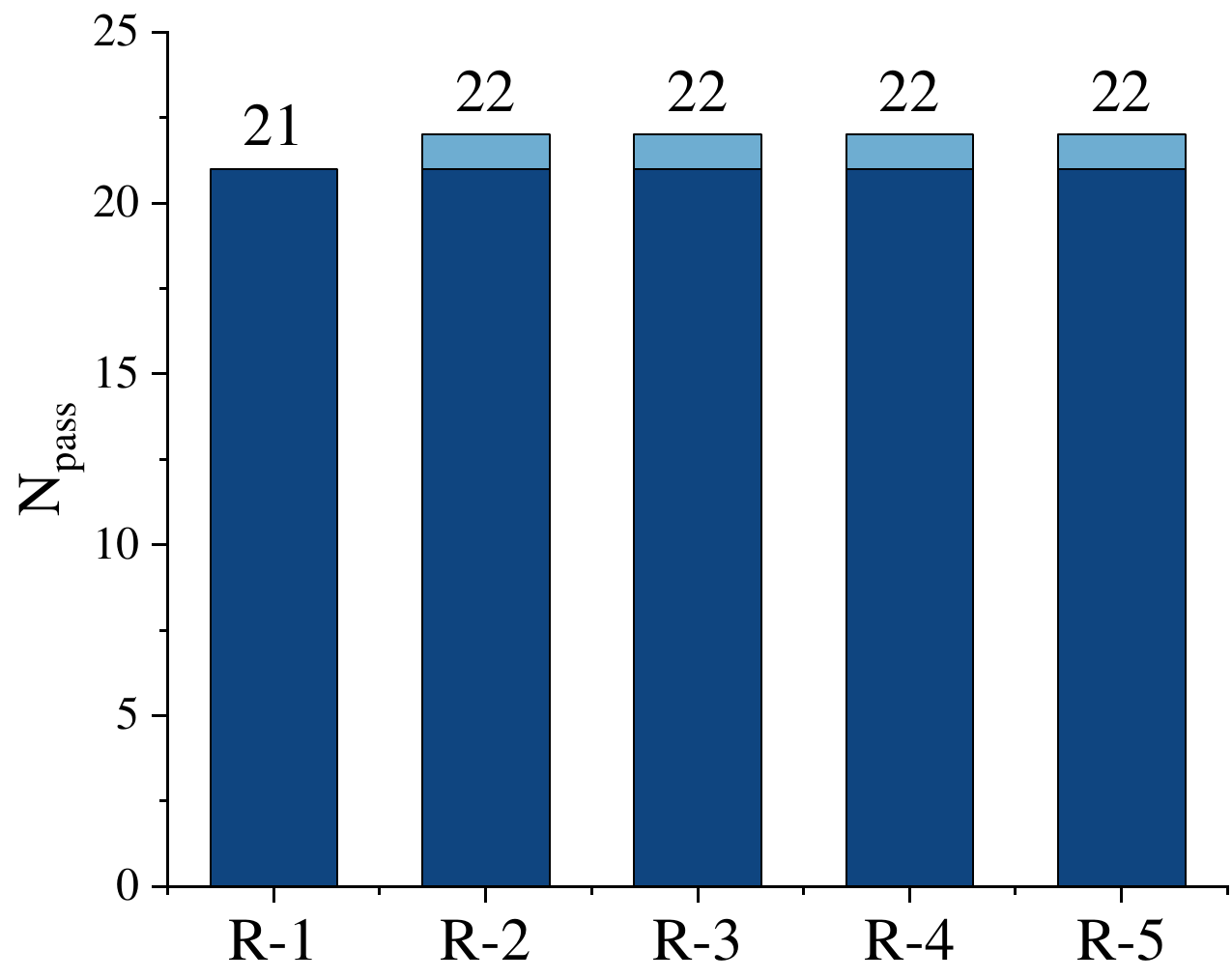}
        \caption{\tiny Codex (Python$\rightarrow$C++)}
        \label{rq3figure7}
    \end{subfigure}
    \hfill
    \begin{subfigure}[b]{0.22\textwidth}
        \centering
        \includegraphics[width=\textwidth]{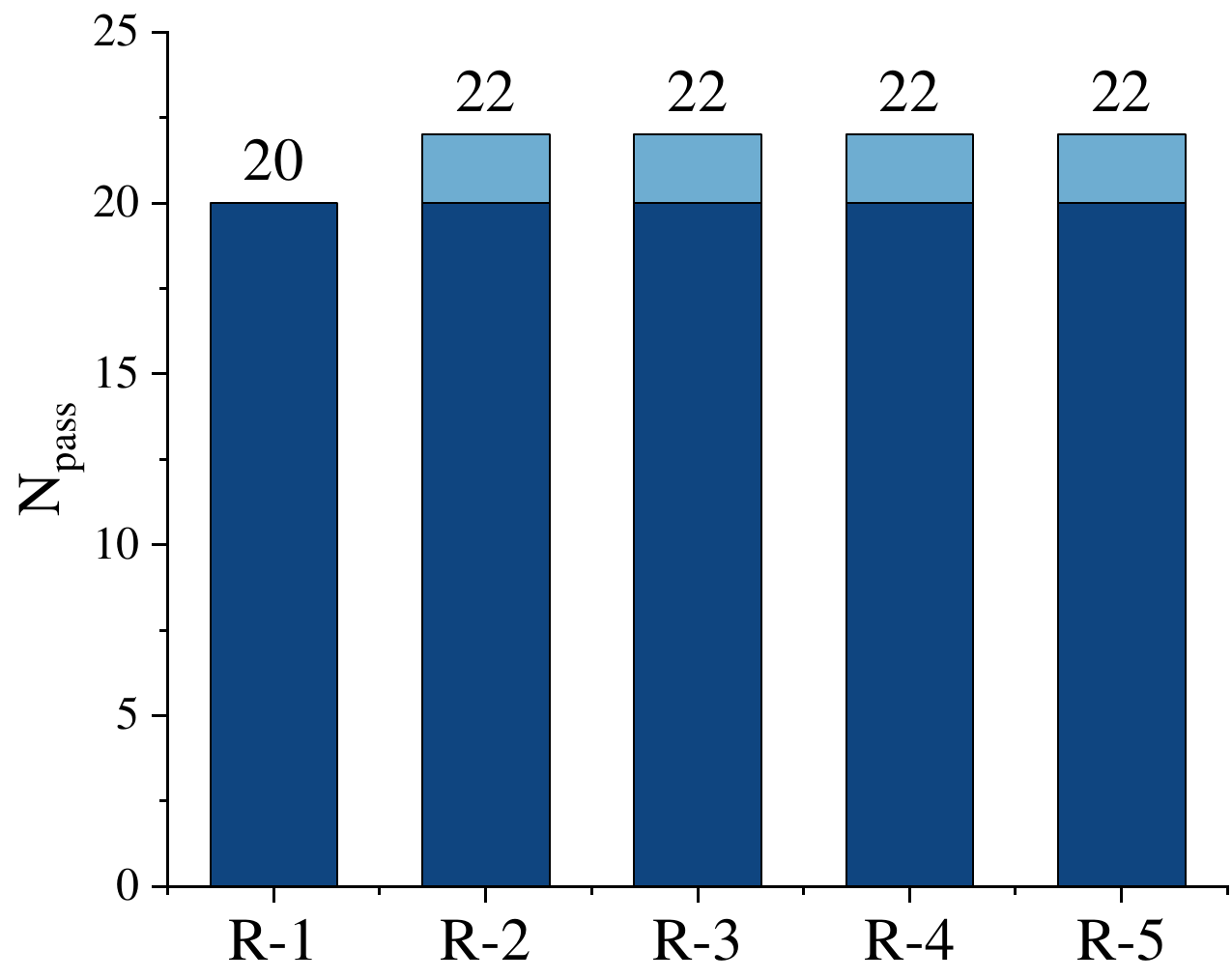}
        \caption{\tiny Qwen2.5 (Python$\rightarrow$C++)}
        \label{rq3figure8}
    \end{subfigure}
\caption{Total number of fixed translations after RulER executes multiple align-locate-repair iterations. ``R-n'' indicates performing the n-th round of align-locate-repair iteration. ``Qwen2.5'' refers to Qwen2.5-Coder-32B-Instruct.}
\label{rq3figure}
\end{figure}

As some translations may contain multiple semantic errors and require multi-line code modifications to be successfully repaired, we further study how many \textit{align-locate-repair} iterations are necessary for RulER to fix the erroneous code translations. Figure~\ref{rq3figure} presents the cumulative number of repaired translations after each align-locate-repair iteration. The results indicate that the majority of successfully repaired translations by RulER are achieved after executing \textit{one or two align-locate-repair iterations}, while some require \textit{three, four, or even five iterations} to be fully repaired. This demonstrates RulER's capability to address complex semantic errors that require multi-line modifications.

\section{Discussion}
\label{sec:discussion}

\subsection{Revisiting the Role of Code Translation Rules in Automated Code Translation}

Automated code translation has progressed from rule-based code transpilers to learning-based code translation models. Early transpilers rely on manually crafted translation rules to translate code, but the generated translations suffer from poor readability and maintainability. To address these limitations, researchers turned to learning-based methods, developing specialized DL models or directly prompting LLMs to generate more flexible and natural translations. Nevertheless, the inherent black-box nature of DL models makes it challenging to guarantee the correctness of their translation outputs.

In this paper, we propose to reintroduce the idea of adopting translation rules to enhance the current learning-based methods. While it seems that using translation rules is no longer the primary choice \textbf{for generating translation} today, they can still offer a relatively rigorous reference to improve the translations generated by DL models. Specifically, we propose adopting translation rules to: (1) construct code alignments to help locate errors in translations, and then (2) generate repair patches to fix these errors. However, existing translation rules designed for transpilers are not suitable for building code alignments across various translations. Due to the high cost of manually designing these rules, transpiler rules typically capture only a rigid one-to-one mapping between the source code and a single fixed translation. In contrast, code translations can be highly flexible and varied, and aligning these translations requires diverse translation rules as references.

\begin{table}[t]
\centering\footnotesize
    \caption{Different C++ translations of Python's exponentiation operator \code{**} and whether they can be covered by the rules of py2many or RulER.}
    \label{diss1table2}
    \begin{tabular}{cllcc}
        \toprule
        ID & Python Code & C++ Translation & py2many & RulER \\
        \midrule
        Case-1 & \texttt{\footnotesize a ** b} & \texttt{\footnotesize std::pow(a, b)} & \ding{51} & \ding{51} \\
        Case-2 & \texttt{\footnotesize a ** 0.5} & \texttt{\footnotesize std::sqrt(a)} & \ding{55} & \ding{51} \\
        Case-3 & \texttt{\footnotesize 2 ** a} & \texttt{\footnotesize 1 << a} & \ding{55} & \ding{51} \\
        Case-4 & \texttt{\footnotesize a ** 2} & \texttt{\footnotesize a * a} & \ding{55} & \ding{51} \\
        Case-5 & \texttt{\footnotesize 10 ** 5} & \texttt{\footnotesize 100000} & \ding{55} & \ding{51} \\
        \bottomrule
    \end{tabular}
\end{table}

To address the limitations of existing rules, we propose an automated method to extract translation rules based on the correct code translations generated by LLMs. These extracted translation rules can handle more diverse translations than those used by transpilers, making them more effective for constructing code alignments across various translations.
Take the exponentiation operator \code{**} in Python as an example. 
For any Python expression containing \code{**}, py2many, a powerful Python transpiler capable of converting Python code into C++, can only translate it into \code{std::pow()} in C++. Actually, as shown in Table~\ref{diss1table2}, there can be several correct C++ translations for Python expressions involving \code{**}. Beyond the basic \code{std::pow()} (Case-1), the translations may include operations like square root (Case-2), bitwise shifts (Case-3), repeated multiplication (Case-4), or constant folding (Case-5). The rules we mined based on LLMs can encompass all the above C++ translations.

In summary, we propose that leveraging DL models to generate code translations and employing translation rules for refining these translations presents a promising direction. Additionally, we propose an automated rule construction method to address the limitations of manually constructed rules, thereby enabling them to more effectively improve the code translations generated by DL models.

\subsection{What types of semantic errors can RulER fix that BatFix cannot?}

\begin{figure}[t]
    \centering
    \begin{subfigure}[b]{0.325\textwidth}
        \centering
        \includegraphics[width=\textwidth]{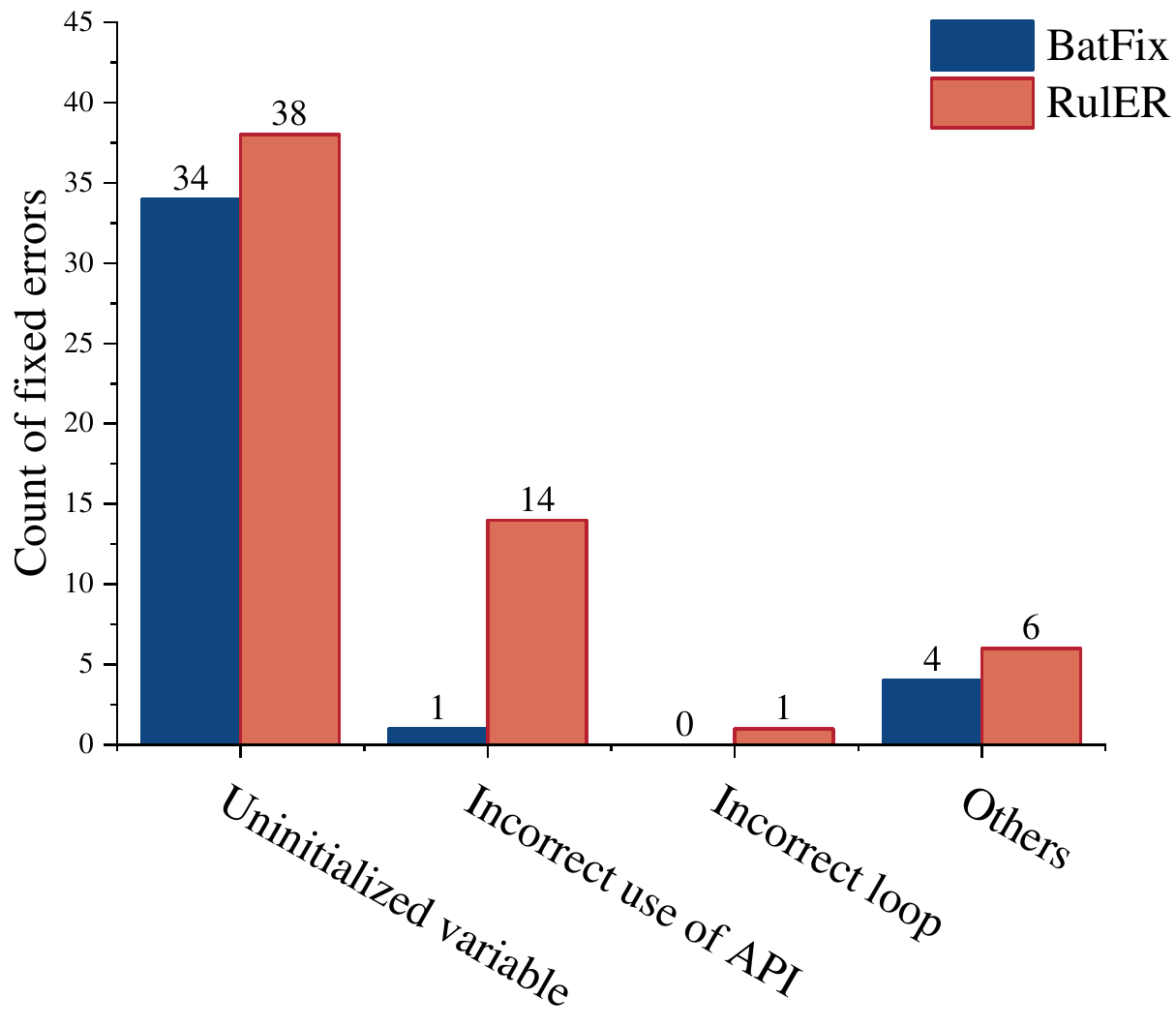}
        \caption{Java$\rightarrow$C++}
        \label{veen1}
    \end{subfigure}
    \hspace{5em}
    \begin{subfigure}[b]{0.45\textwidth}
        \centering
        \includegraphics[width=\textwidth]{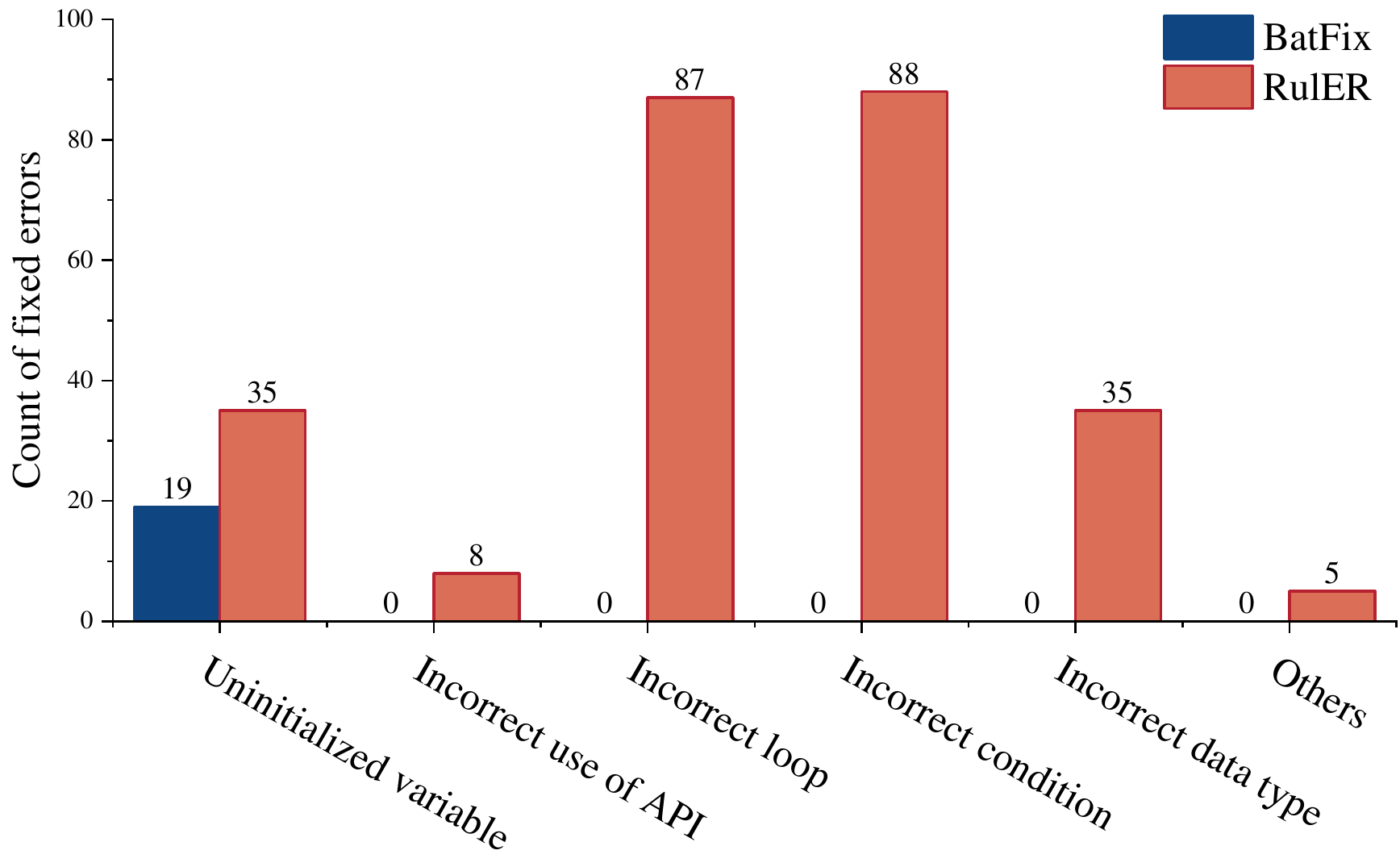}
        \caption{Python$\rightarrow$C++}
        \label{veen2}
    \end{subfigure}
\caption{Types and counts of semantic errors fixed by BatFix and RulER in Java-to-C++ and Python-to-C++ translations. The sum of fixed errors may exceed the number of translation programs repaired by the method, as a single translation program can contain multiple errors.}
\label{fix-types}
\end{figure}

To gain deeper insights into the semantic errors that RulER can fix but BatFix fails to resolve, we manually annotated the types of semantic errors each method corrected in the evaluation dataset. Figure~\ref{fix-types} presents a statistical comparison of the semantic errors fixed by RulER and BatFix across different error types.

\begin{figure*}[t]
\centering
\includegraphics[width=1.0\linewidth]{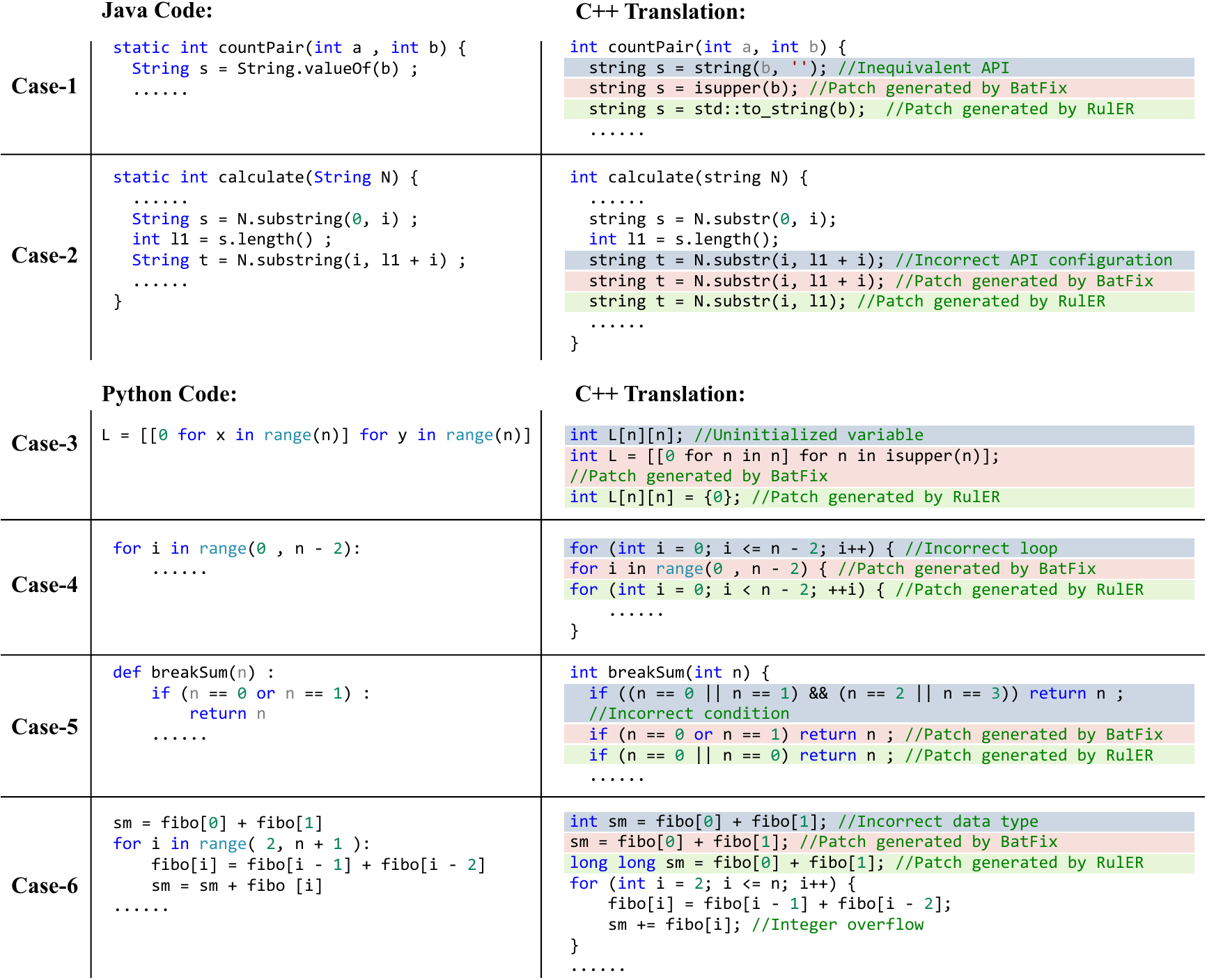}
\caption{Examples of semantic errors that RulER can resolve but BatFix fails to fix. The code highlighted in \textcolor{RoyalBlue}{blue} represents the original erroneous translation, the code highlighted in \textcolor{red!70}{red} is the incorrect repair patch generated by BatFix, and the code highlighted in \textcolor{LimeGreen}{green} shows the correct repair patch produced by RulER.}
\label{fix-examples}
\end{figure*}

For Java-to-C++ translations, RulER significantly outperforms BatFix in fixing ``Incorrect use of API'' errors, which involve using inequivalent APIs or misconfiguring parameters. For these errors, BatFix lacks references to identify the equivalent API and tends to replicate the API parameter configurations from the source Java code into C++ repair patches, which often leads to incorrect API usage due to the differences between the Java and C++ APIs. 
Figure~\ref{fix-examples} illustrates two examples, Case-1 and Case-2, where BatFix produces repair patches with either inequivalent APIs or incorrect parameter configurations. In Case-1, the source Java code uses \code{String.valueOf()} to convert variable \codeWoQuot{b} into its string representation, but the generated C++ translation incorrectly employs the constructor \code{string()}, which cannot convert a variable into a string. Worse still, the repair patch generated by BatFix invokes an entirely irrelevant API \code{isupper()}. In Case-2, the Java API \code{substring()} specifies the end position of the substring as its second parameter. In the generated C++ translation, although \code{substr()} is correctly chosen as the corresponding API, its second parameter, which represents the length of the substring, is mistakenly set to the end position of the substring, i.e., \code{l1+i}. However, BatFix replicates both errors in its repair patches due to mimicking the structure of the original Java code.
These examples reveal that BatFix fails to address ``Incorrect use of API'' errors due to its lack of guidance on selecting appropriate APIs in the target PL and configuring their parameters correctly. In contrast, RulER leverages code translation rules, which include mappings for various API translations. These rules enable RulER to identify equivalent APIs in the target PL (e.g., mapping Java's \code{String.valueOf()} to C++'s \code{std::to\_string()}) and determine the correct parameter configurations (e.g., mapping Java's \code{substring(param1, param2+param1)} to C++'s \code{substr(param1, param2)}). With these API mappings, RulER proves more effective than BatFix at repairing ``Incorrect use of API'' errors.

For Python-to-C++ translations, RulER exhibits a clear edge over BatFix in addressing errors of ``Incorrect loop'', ``Incorrect condition'', and ``Incorrect data type''. These errors are challenging to fix for BatFix because the expected C++ patches often exhibit different structures or contain different identifiers from the source Python code. 
Specifically, 
(1) Python implements for-loops using \code{range()}, whereas C++ achieves this through initialization, iteration, and conditional checks.
(2) Logical operators differ between Python and C++, such as Python's \code{and} and \code{or}, which correspond to \code{\&\&} and \code{||} in C++.
(3) Python relies on automatic type inference during variable assignment without explicit type declarations, while C++ requires explicit type specifications.
As a result, the repair patches generated by BatFix referencing the source Python code often have issues such as incorrect loop structures, the inclusion of Python-specific logical operators, or missing type declarations, as illustrated by Case-4, Case-5, and Case-6 in Figure~\ref{fix-examples} respectively. 
In summary, BatFix fails to fix these errors because it relies solely on the source Python code when generating repair patches for C++, making it difficult to satisfy the grammar requirements in C++.
In contrast, RulER generates repair patches based on translation rules, allowing it to further infer the expected structures of translated C++ code from the source Python code, including the traditional C++ for-loop structures, corresponding logical operators (\code{\&\&} and \code{||}), and explicit type declarations. Leveraging the rule-derived code structures, RulER can generate patches that align more closely with C++ grammar requirements. As a result, RulER outperforms BatFix in fixing errors related to ``Incorrect loop'', ``Incorrect condition'', and ``Incorrect data type''.

\section{Threats to Validity}
\label{sec:threatstovalidity}

One threat to validity arises from the representativeness of the evaluation dataset. To alleviate this threat, we use an existing evaluation dataset collected by the baseline work~\cite{BatFix} and further expand the dataset to enable a more comprehensive assessment. Specifically, the dataset used in our evaluation consists of 553 erroneous code translations for diverse algorithms. These translations are generated by four representative code translation models across two different PLs, namely Java-to-C++ and Python-to-C++, ensuring well representativeness.

Another threat to validity is the subjective judgment of the annotators when labeling code alignments and buggy code lines within the evaluation dataset. To mitigate this threat, we employed two graduate students experienced in Java, C++, and Python programming to independently annotate the dataset. The Cohen's Kappa coefficient calculated from their annotations, i.e., 0.839 for code alignment annotations and 0.897 for semantic error localization annotations, indicated a substantial level of agreement. Finally, any discrepancies between their annotations were thoroughly discussed and resolved through consensus, ensuring the reliability of the annotated labels.

The last threat to validity is whether RulER can adapt to various PLs. To address this threat, we evaluate RulER in locating and repairing semantic errors using code translations across two language pairs, i.e., Java-to-C++ and Python-to-C++. The experimental results confirm that RulER exhibits good generalizability across both PL pairs. Moreover, RulER should be able to be easily adapted to a wide range of PLs. The LLM employed by RulER to generate code translations, i.e., Qwen2.5-Coder-32B-Instruct, is capable of handling translation tasks across numerous PLs due to its training data encompassing nearly 40 PLs~\cite{qwen2.5}.
This capability enables RulER to mine translation rules across various PLs. As a result, RulER can be readily adapted to diverse PL pairs with minimal adjustments required.

\section{Related Work}
\label{sec:relatedwork}

\subsection{Fault Localization}

Fault Localization (FL) has been a widely studied task in software engineering and serves as a critical step in automated program repair~\cite{FL-survey1}. Existing studies have proposed various FL techniques, which can be categorized into slice-based~\cite{FL-slice1, FL-slice2, FL-slice3}, program spectrum-based~\cite{FL-spectrum1, FL-spectrum2, FL-spectrum3}, statistics-based~\cite{FL-statistic1, FL-statistic2, FL-statistic3, FL-statistic4}, program state-based~\cite{FL-state1, FL-state2, FL-state3}, and learning-based approaches~\cite{FL-DL1, FL-DL2, FL-DL3, FL-DL4, FL-DL5, FL-DL6}. These FL techniques are designed to locate faults within programs written in a single PL. In contrast, the FL for semantic errors in code translation aims to identify the semantic discrepancies between programs written in two different PLs. TransMap was the first to propose locating semantic errors in code translations through code alignment. TransMap leverages an LLM to directly generate code alignments between the source and target programs, based on which the semantic errors can be located by comparing the execution traces and variable values of aligned statements. Subsequently, BatFix introduced another code alignment algorithm based on control flow graph analysis. 
However, the references used by these methods for code alignment are not reliable in certain translation scenarios. To address this limitation, in this paper, we propose leveraging code translation rules to provide more reliable guidance for code alignment and introduce an automated method for constructing code translation rules to mitigate the difficulty of manually building such rules.

\subsection{Automated Program Repair}

Automated Program Repair (APR) aims to automatically fix defects in buggy programs and reduce manual debugging work. Most APR techniques follow a generic repair framework consisting of fault localization, patch generation, and patch validation~\cite{APR-survey1}. 
Among these steps, patch generation has attracted much attention from researchers. Various approaches have been proposed for generating repair patches, which can be broadly categorized into search-based~\cite{APR-search1, APR-search2, APR-search3}, constraint-based~\cite{APR-constraint1, APR-constraint2}, template-based~\cite{APR-template1, APR-template2}, and learning-based methods~\cite{APR-learning1, APR-learning2, APR-learning3, APR-LLM1, APR-LLM2, APR-LLM3}. Unlike these, the APR of code translations benefits from the existence of the source program as a reference solution. Therefore, BatFix proposes mimicking the structure of the source program to design templates for synthesizing repair patches. However, BatFix performs well only when the source and target PLs are highly similar in syntax, but struggles if they differ significantly. To address this limitation, we propose using code translation rules to further infer the expected structures of the correct translations, thereby guiding the design of more accurate templates to generate more effective repair patches.

\section{Conclusion and Future Work}
\label{sec:conclusion}

Existing methods for diagnosing buggy code translations rely on code alignment to locate and fix semantic errors. 
However, these methods lack reliable references for constructing code alignments and designing repair patch generation templates, limiting their effectiveness in locating and fixing semantic errors in code translation.
To address these limitations, we propose a rule-based code translation debugging method called RulER, which leverages code translation rules to build code alignments and design patch generation templates. Translation rules capture clear and detailed structural correspondences between different Pls, enabling RulER to construct more precise code alignments and facilitate the derivation of more accurate templates for patch generation. 
To mitigate the high cost of manually constructing code translation rules, RulER employs a removal-based strategy to mine translation rules from correct translation results generated by LLM. Additionally, it can combine the mined rules to construct new ones, further enriching its rule set.
Based on the rules, RulER outperforms two state-of-the-art methods, BatFix and TransMap, in locating and repairing semantic errors in code translation. In our evaluation, RulER achieves relative improvements of 20\% and 272\% over the best baseline in terms of success rates for error localization and repair, respectively. Furthermore, the evaluation results indicate that generating repair patches based on the rules mined from LLM is more effective than directly prompting LLMs to fix the erroneous translations.
In the future, we plan to run RulER to mine translation rules across a wider range of PLs and share the rules as reusable materials for later code translation research. We will also explore the potential of these rules to enhance the learning-based code translation models by integrating them into the training process.

\section*{Acknowledgments}
This work was supported by National Key R\&D Plan of China (Grant No.2024YFF0908003), National Natural Science Foundation of China (No. 62472326), CCF-Zhipu Large Model Innovation Fund (No. CCF-Zhipu202408), and Hong Kong Research Grant Council/General Research Fund (No. 16206524).

\clearpage
\end{CJK*}
\bibliographystyle{ACM-Reference-Format}
\bibliography{ref}

\end{document}